\documentclass[jgrga]{agutex}

  \usepackage[dvips]{graphicx}
  \setkeys{Gin}{draft=false}
\authorrunninghead{GIORDANO ET AL.}

\titlerunninghead{UVCS/SOHO CME CATALOG}

\authoraddr{Corresponding author: S. Giordano,
INAF/Osservatorio Astrofisico di Torino
via Osservatorio 201, Pino Torinese, 10025 Italy
(giordano@oato.inaf.it)}

\begin{document}

\title{UVCS/SoHO Catalog of Coronal Mass Ejections from 1996 to 2005: Spectroscopic Proprieties  }

\authors{S. Giordano\altaffilmark{1}, 
	A. Ciaravella\altaffilmark{2}, 
	J.C. Raymond\altaffilmark{3}, 
	Y.-K. Ko\altaffilmark{4},  and 
	R. Suleiman\altaffilmark{3}}
\altaffiltext{1}{INAF-Osservatorio Astrofisico di Torino, via Osservatorio 20, 10025 Pino Torinese, Italy} 
\altaffiltext{2}{INAF-Osservatorio Astronomico di Palermo, P.za Parlamento 1, 90134 Palermo, Italy}
\altaffiltext{3}{Harvard-Smithsonian Center for Astrophysics, 60 Garden St., Cambridge, MA 02138, USA}
\altaffiltext{4}{Space Science Division, Naval Research Laboratory, Washington, DC 20375, USA}

\begin{abstract}
Ultraviolet spectra of the extended solar corona have been routinely obtained by SoHO/UVCS since 1996. 
Sudden variations of spectral parameters are mainly due to the detection of Coronal Mass Ejections (CMEs) 
crossing the instrumental slit. We present a catalog of CME ultraviolet spectra based upon a systematic search 
of events in the LASCO CME catalog, and we discuss their statistical properties.
Our catalog includes 1059 events through the end of 2005, covering nearly a full solar cycle. It is online 
available at the URL $\rm http://solarweb.oato.inaf.it/UVCS\_CME$ and embedded in the online LASCO CME 
catalog ($\rm http://cdaw.gsfc.nasa.gov/CME\_list$).
The emission lines observed provide diagnostics of CME plasma parameters, such as the light-of-sight velocity, 
density and temperature and allow to link the CME onset data to the extended corona white-light images. 
The catalog indicates whether there are clear signatures of features such as shock waves, current sheets, 
O~VI flares, helical motions and which part of the CME structures (front, cavity or prominence material) are 
detected. The most common detected structure is the cool prominence material (in about 70\% of the events).
For each event, the catalog also contains movie, images, plots and information relevant to address detailed 
scientific investigations. The number of events detected in UV is about 1/10 of the LASCO CMEs, and about 1/4 of 
the halo events. We find that UVCS tends to detect faster, more massive and energetic CME than LASCO and 
for about 40\% of the events events it has been  possible to determine the plasma light-of-sight velocity.

\end{abstract}

\begin{article}


\section{Introduction}\label{intro}

Coronal Mass Ejections (CMEs) are dramatic eruptions of magnetized plasma from the Sun into the interplanetary medium
observed for the first time 
from space 
in 1971 by the Orbiting Solar Observatory (OSO) coronagraph \citep{Tou73}.
Over the next two decades, CMEs were detected by white-light spaceborne coronagraphs aboard Skylab, P87-1 and Solar Maximum 
Mission (SMM) and ground-based K-Coronameter at Mauna Loa. 
Only few ultraviolet spectra of coronal transient events were available before the Solar and Heliospheric Observatory (SoHO) mission, 
in particular a couple of prominence eruptions observed in the low corona (below $\sim$1.5R$_{\odot}$) by SO55 experiment aboard Skylab 
\citep{Sch77} and by UVSP experiment aboard SMM \citep{Fon89}.
Since the launch of the SoHO the Large Angle and Spectrometric Coronagraph (LASCO) telescope routinely observed CMEs in 
white-light images, increasing by about an order of magnitude the number of events previously known and for the first time ultraviolet 
spectroscopic observations of CMEs have been obtained with the UltraViolet Coronagraph Spectrometer (UVCS) and at lower heights by 
the 
Solar Ultraviolet Measurements of Emitted Radiation (SUMER) 
and 
Coronal Diagnostic Spectrometer (CDS) 
spectrographs
telescopes. 
These spectroscopic observations have been an important link between the EUV 
and X-ray
imaging of the CME onset obtained with EUV Imaging 
Telescope (EIT), Transition Region and Coronal Explorer (TRACE) or Hinode/X-Ray Telescope (XRT) 
and the white-light images 
by 
LASCO C2 and C3 coronagraphs (e.g. \citet{Gal03, Man07, Lan10}). 
UV spectra provide important diagnostics of the CME plasma, and in many observations the UV spectra overlap with the 
white-light images, allowing further diagnostics of the CME plasma.

Doppler shifts of the CME front and prominence core have shown very large line of sight (LOS) speeds in many events, 
including some originating at the limb. 
The LOS speed combined with the LASCO plane-of-sky speed can be used to determine the total speed and the angle with 
the plane of the sky. The latter has been used to compute the true height of the observed portion of the CME \citep{Cia05,Cia06}.  
In this way, Doppler shift measurements from UVCS are needed for the 3D reconstruction of the true CME morphology and 
speed \citep{Cia00,Lee06,Lee09} and to identify CMEs originating from the backside of solar limb \citep{mie07}. 
Sometimes changes in the Doppler shift with time suggest helical motion \citep{Ant97, Cia97, Cia00, Cia01}.

An independent diagnostic for the total (outflow) speed of a CME comes from the ratio between lines with collisional and radiative 
components, such as the O~VI \mbox{$\:\lambda\lambda$}1032, 1037 doublet.
The radiative component originates from the resonant scattering of the chromospheric line by oxygen ions of the corona. 
As the ions move outward in the corona, the radiative component dims because the solar emission and coronal absorption profiles 
are Doppler shifted apart. Nearby lines of C II \mbox{$\:\lambda\lambda$}1036.34, 1037.02 can pump the radiative component of 
the $\lambda$1037 line at outflow speeds of 170 and 370~km~s$^{-1}$,  respectively \citep{Li98, Dod98}.
In the 2000 June 28 CME \citep{Ray04} obtained outflow speeds in different regions of 
1650~km~s$^{-1}$ and 1810~km~$s^{-1}$ from the detection of pumping of the $\lambda$1037 line
by $\lambda$1032 and $\lambda$1032 line by H I Ly$\beta$ $\lambda$1026 respectively.

Density diagnostics for CMEs have been obtained from O~VI doublet ratios or from density sensitive line ratios 
such as O~V \mbox{$\:\lambda\lambda$}1213.85, 1218.39 \citep{Cia99,Akm01,Ray04,Lee09,Mur11}.
It is also possible to obtain the density by dividing the emission measure, ${n_e}^2L$, derived from emission line intensities, 
by the column density, ${n_e}L$, which is derived from white-light coronagraph images  \citep{Cia08,Lee09,Mur11}.
The combination of density and ionization state makes it possible to infer the heating rate in the ejected plasma 
\citep{Akm01, Lee09, Lan10, Mur11}.
In some cases the number of simultaneously observed spectral lines allows to determine the elemental composition 
of the ejected plasma and to infer where the CME material originated \citep{Cia97,Cia02,Cia03}. 

The analysis of line profiles provides diagnostics of several properties of the detected plasma, such as temperature, 
bulk expansion or shocks. 
Both expansion of the emitting volume and increasing temperature contribute
to broadening the line profiles \citep{Ray00,Cia97,Cia05}.
The heating due to the passage of a shock causes broad wings mainly 
in the ions. The neutral atoms, like the hydrogen, are not 
directly affected by an MHD
collisionless shock, but they react on a longer timescale following charge transfer 
with protons and ionization by electrons.  Detections of broad O~VI line wings have
shown the presence of shocks in several CME fronts \citep{Ray00,Man02,Rao04,Cia05,Cia06,Man08,Bem10,Man11}.
 
Evidence for current sheets (CS) in the wake of CMEs has been detected in the UV spectra as spatially very narrow emission in the 
hot (6 $\times$ 10$^6$ K) line of Fe~XVIII \citep{Cia02,Ray03,Ko03,Bem06,Cia08,Sch09}, supporting models that emphasize the role 
of the reconnection current sheet connecting the detaching flux rope to the post-CME arches in the dynamics of the CME \citep{Lin00}.
UVCS observations have been used also to test the \cite{Vrs09} model of post-CME CS \citep{Ko10} by comparing calculated and 
observed physical properties inside the post-CME CS.
Magnetic reconnection layers induced by plasma expansion of slow CME has been identified also by comparing structures detected in 
white-light images from LASCO and in UV radiation from UVCS \citep{Bem08,Pol08}.
Observations of narrow bright feature in the Fe~XVIII which provides estimated of CS thickness and temperatures far larger that 
expected \citep{Cia08} could be explained by plasma turbulent reconnection \citep{Bem08b}.

CMEs are most often seen in lines of cool ions such as O~VI, H I Ly$\alpha$ and C~III. 
This is erupting prominence material that is denser than its surroundings. 
It often takes the form of long filaments, presumably tracing out magnetic field structures. 
The bright prominence cores have been detected in the 3 part structured CMEs; 
the bright core inside a dark void surrounded by the CME front. 
The observed bright filament cores exhibit the most variable Doppler shift signatures in UVCS data.
Voids are usually detected as depletion in the background corona emission although sometime some weak diffuse 
emission from hot lines such as Si XII has been observed \citep{Cia03}.  
The leading edge is usually seen as a modest brightening of the usual coronal lines.

This paper presents the results of a systematic search for CMEs in archival UVCS data.  
We have created a catalog of 
1059 events mainly based on the CDAW catalog of CMEs observed 
by LASCO between 1996 and 2005 \citep{Yas04,Gop09}. 
The catalog entry for each event includes a movie of LASCO difference images with the UVCS slit positions overlaid, 
a list of detected emission lines, maximum and minimum Doppler velocities, and 2D images of line intensities 
as functions of polar angle and time.  It also indicates whether signatures of CME features such as shock fronts, 
current sheets or prominence material are present.
The catalog more than triples the number of known UVCS observations of current sheets and of 
detections of O~VI emission from flares scattered by coronal ions \citep{Joh11}.  
We discuss the selection criteria for CMEs in the UVCS data and the criteria for identifying physical features 
such as shock waves and current sheets. 
The catalog, here briefly described, is available online to the community hosted by the Astrophysical Observatory of 
Torino and also included into CDAW LASCO catalog, moreover, a search tool is provided to select events which meet 
user-given parameters.
Following the description of the catalog, we examine the statistics of CMEs observed by UVCS, 
such as CME frequency during the solar cycle, and compare them with the white-light properties from LASCO.  
In general, UVCS sees about one tenth of the LASCO CMEs, with a bias toward the more massive,
more energetic and wider events.  The Doppler shifts are on average about 1/10 of the speeds of the leading edges 
from the CDAW catalog, both because most UVCS events occur near the limb and because UVCS often observes 
the slower prominence material.  
Finally, we present a summary of the observed intensities from the main detected spectral lines which can be
useful for designing new  instruments to observe CMEs in the UV wavelength range.

\section{UVCS Observation of CMEs}\label{cme}

UVCS is a slit spectrometer designed for observation of the solar corona from 1.4 up to 10~R$_{\odot}$. 
The instrumental slit,  40$^\prime$ long, is perpendicular to the sunward direction on the plane of the sky, 
and its width can be varied from few arcseconds up to 84$^\prime$$^\prime$.
It can be rotated by 360$^{\circ}$ to cover the full corona.  
The spectrometer has two UV channels, LYA and OVI, optimized to detect the H~I Ly$\alpha$ 1216\AA\/ line and 
the O~VI \mbox{$\:\lambda\lambda$}1032, 1037 
doublet, respectively.  They cover from 945  to 1270 \AA\/ in first order and  from 473 to 635 \AA\/ in second order 
(see \citet{kohl95} for a detailed description of the instrument). The OVI channel also includes a 
redundant path that focuses the H I Ly$\alpha$ line onto the detector.
The spectra are two dimensional images of 360 $\times$ 1024 pixels giving the line intensity 
distribution along the slit as a function of the wavelength. The spatial pixel is 7$^\prime$$^\prime$ while 
spectral pixels are 0.0993 \AA\ (0.0915 \AA\ for the redundant path) and 0.1437\AA\/ for OVI and LYA channels,
respectively.  Due to telemetry limitations the detectors can be masked to select the wavelength ranges 
of interest with the desired spectral and spatial binning. The full detector can be obtained with a 
spectral binning of 2 pixels and a spatial of 10 pixels, but most observations  trade some spectral coverage 
for spatial resolution of 3 or 6 pixels.  The UVCS observations of CMEs include a wide range of UVCS 
instrument parameters, as most CMEs were detected during observations intended for other purposes.
Therefore exposure time varies between from 120 sec during a CME watch to 600 sec during coronal hole 
observations. The CME watch programs maintain one position for hours or switch between two 
heliocentric heights. Most of CMEs were observed during the daily synoptic scans, that lasted about 
12h and scanned the corona at eight polar angles (PA) and several heliocentric heights.
 
\begin{figure}
   \centering
	\includegraphics[angle=90,width=7.5cm]{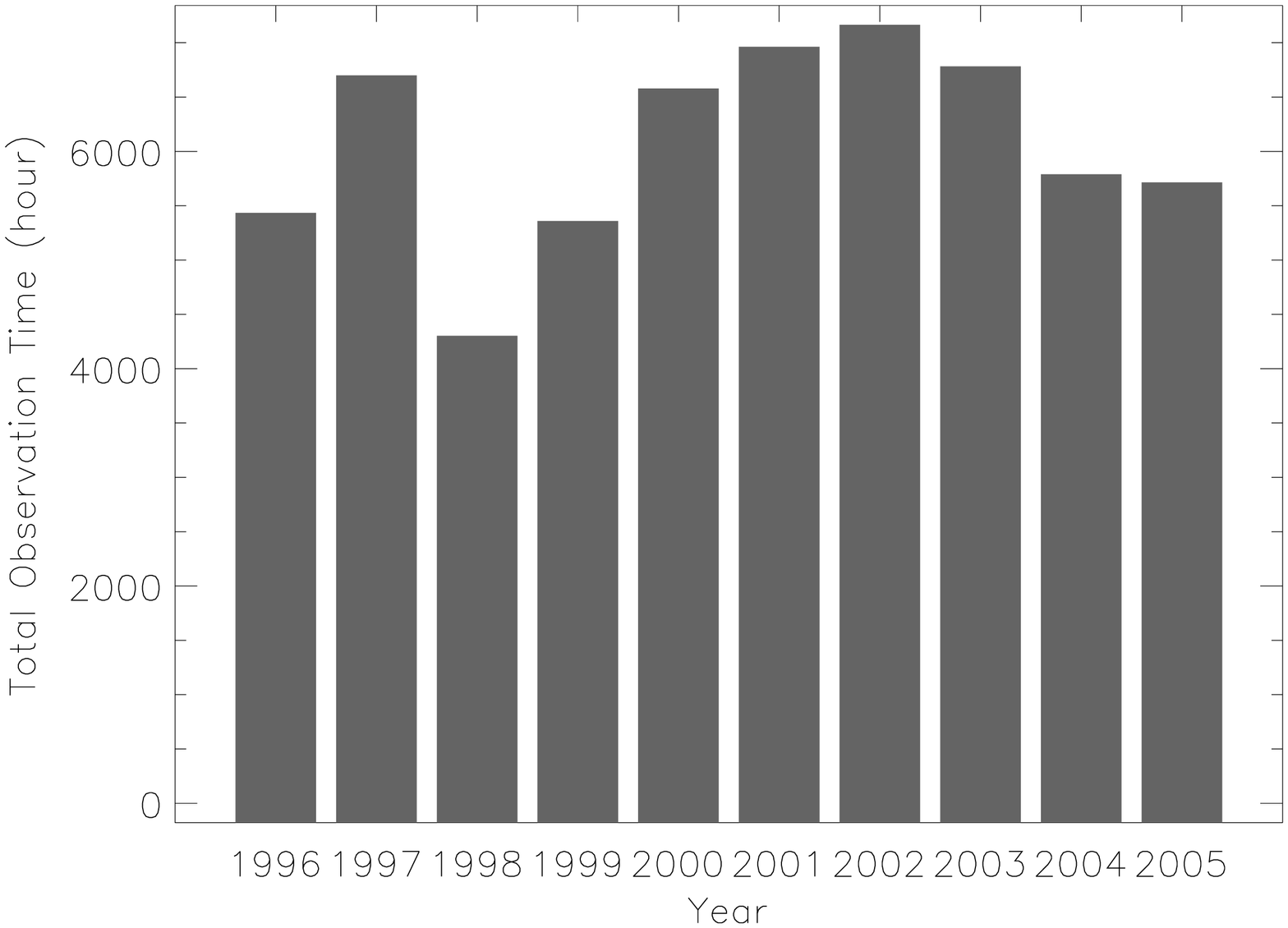} \\
	\includegraphics[angle=90,width=7.5cm]{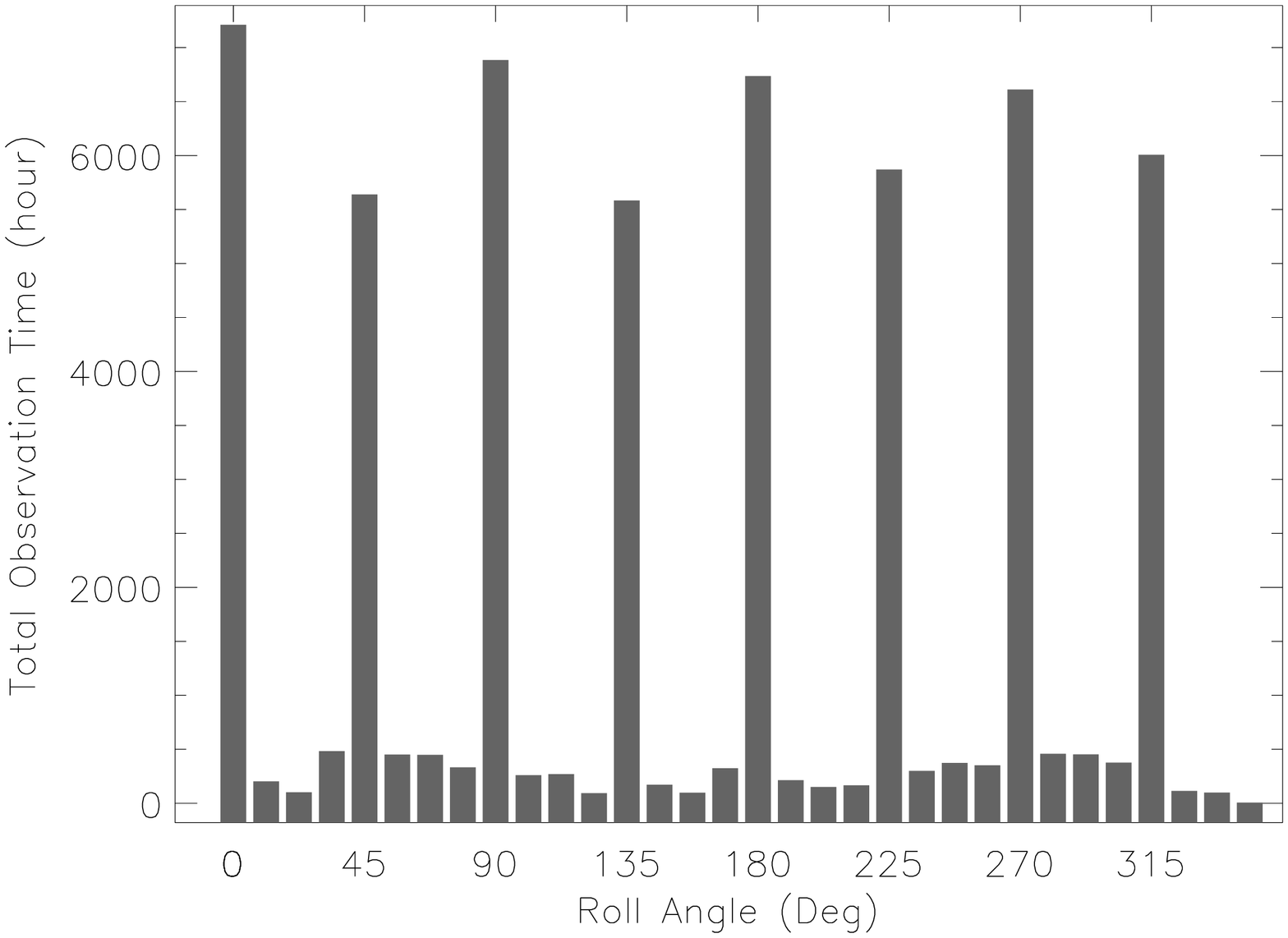} \\
	\includegraphics[angle=90,width=7.5cm]{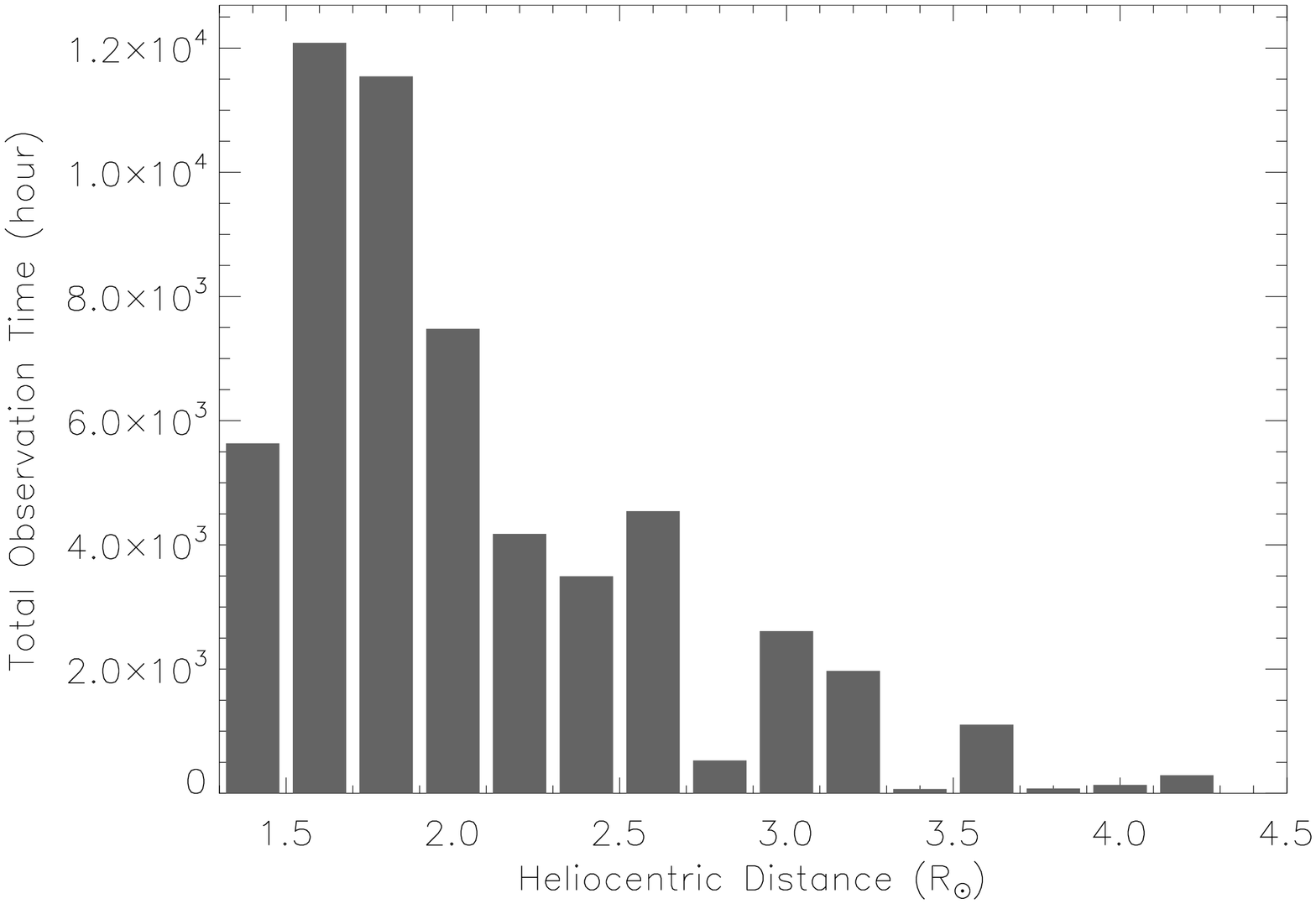}  \\
  \caption{UVCS observations summary from 1996 April 1 to 2005 December 31. 
  		The top panel shows the total observation time per year,
		  the center panel the observation time as a function of the instrumental roll angle and 
		  the bottom panel the distribution of observations as a function of heliocentric height.}
\label{obs_summary}
\end{figure}

The observation period included in the CME catalog covers most of the solar cycle. It starts at 
the beginning of the UVCS science observations, April 1st 1996, near solar minimum and goes through the
end of 2005, well into the declining phase. 
The total exposure time of $\sim 6.1\cdot 10^4$ s, corresponds to a time coverage of about 
71\%, distributed over different roll angles and 
altitudes.  
Figure~\ref{obs_summary} shows the UVCS observation time as functions of year, roll angle 
and heliocentric distance during the selected period. The lower values of observation time 
during 1998 and 1999 are due to periods in which SoHO was not observing. 
The middle panel of Figure~\ref{obs_summary} shows that the longest observation times are at the eight polar angles 
used during daily synoptic program, which scans the full corona with 45$^{\circ}$ angular step. 
Although UVCS also observed at very high heights, especially during comet or star observations, most of the 
observation time was spent below 3.1~R$_{\odot}$ with a maximum at 1.7~R$_{\odot}$, see bottom panel of Figure~\ref{obs_summary}.  
A general review of the CME scientific results obtained with the first decade of UVCS observations 
has been presented by \citet{Kohl06}.

\section{UVCS CME Catalog}\label{uvcs}

UVCS operators and lead scientists through visual inspection of spectral quick look data compiled a preliminary 
list of about 300 CMEs detected by UVCS, we used this list as a guide for implementing a systematic search through the entire 
UVCS observations catalog, embedded in the UVCS data analysis software available as Interactive Data Language (IDL) routines 
in solarsoft \citep{fre98}.
Our data selection is driven by the SoHO/LASCO CME catalog available at the CDAW Data Center 
($\rm http://cdaw.gsfc.nasa.gov/CME\_list$; \citet{Gop09}). For each event in the LASCO CME catalog we searched the UVCS data 
in a time window covering from one hour prior to the CME onset time, estimated from LASCO C2 coronal observations, 
to 4 hours after the time when the CME is expected to leave the UVCS field of view,
(in some case this time window has been extended to include long lifetime CME-related features, such as the current sheets).
Then we select observations pointing within the latitudinal region spanned by the expanding plasma, determined by the angular width 
parameter in the LASCO catalog.
Once the UVCS data potentially containing a CME observation have been selected, a specially designed software tool 
was used to look for temporal changes in line strengths, line widths and line centroids associated with the CME.  
The software produced height-time intensity, width, centroid and running difference maps 
which made it easy to detect changes in any chosen spectral line.
When temporal variations of a spectral parameter are identified, the event is listed as a UVCS CME in the catalog and
the software allows to create an hyper-textual page with embedded a LASCO movie of the event with the UVCS slit 
positions superposed. 
For each UVCS CME identified, by using the NGDC/NOAA events catalogs 
(available at the URL: $\rm ftp://ftp.ngdc.noaa.gov/STP/SOLAR\_DATA$), 
we checked, in the time spent by the CME to expand into solar corona, the detection of a type II radio burst, signature of CME-driven shock, 
and the detection of 
X-ray flares 
in a short time window around the CME onset.
Then the likely association between CME in UVCS catalog and type II radio burst and 
X-ray
flare 
is recorded into the single event pages.
More than 98\% of the CMEs
identified in UVCS data have been detected also by LASCO, therefore, 
they are recorded with a same identification number which allow to link each other the two catalogs.
However, a small number of CMEs without correspondence to a LASCO event were discovered in UVCS data 
(see Table~\ref{uvcsonly}) and included in our catalog. 
As with the LASCO catalog, the UVCS CME catalog is a living document, which can have some minor revisions in the future 
to include new events discovered or in case of more detailed analysis of known events which can provide new interpretation 
of observed features.

\begin{figure}
   \centering
   \includegraphics[width=13cm]{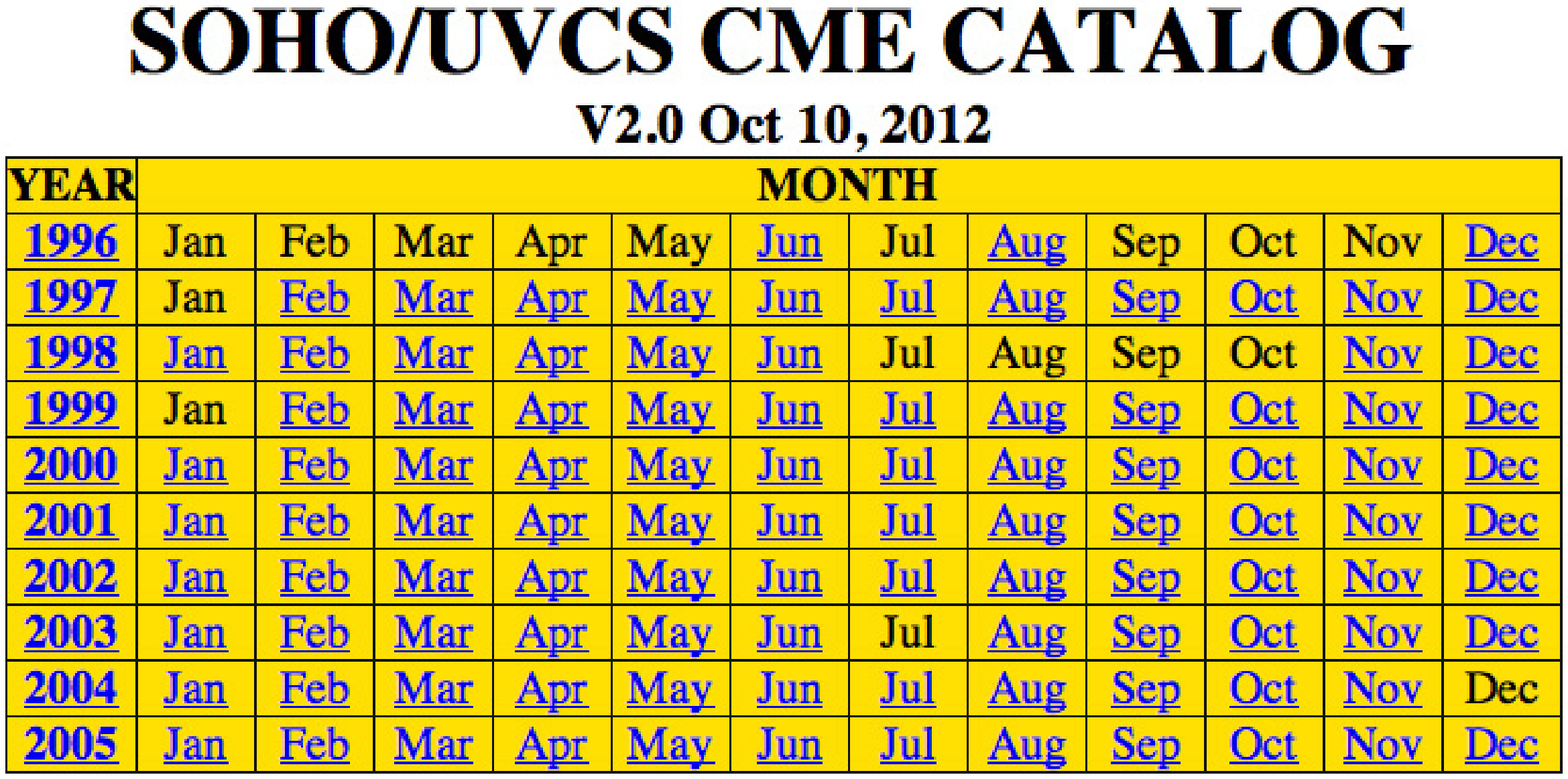}
   \includegraphics[width=17cm]{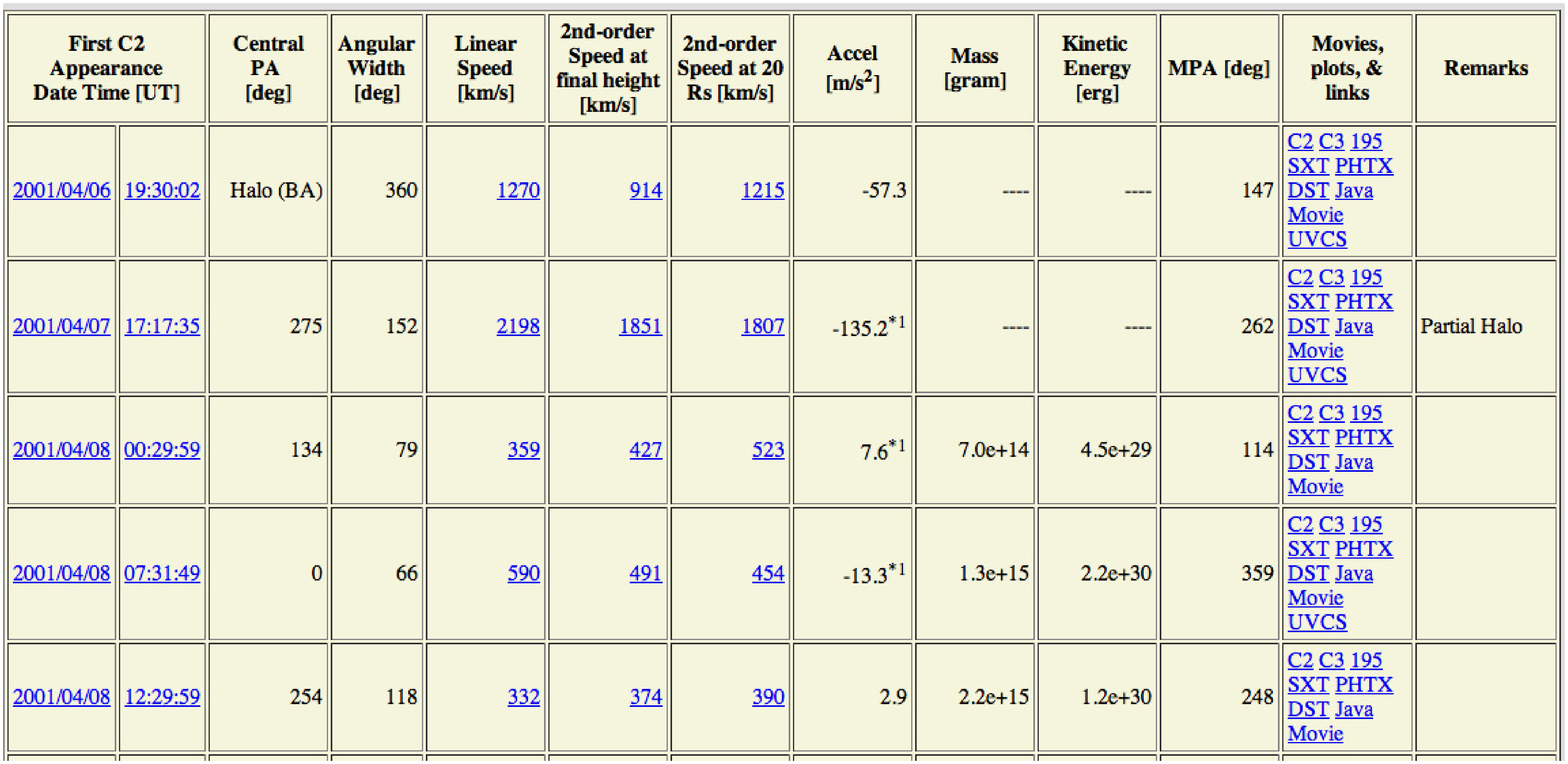}
   \caption{Top panel: Main page of online SoHO/UVCS CME Catalog as a matrix of year and months of observation.
   		Bottom panel: Example of the CME monthly list available at the SoHO/LASCO CME Catalog web pages. 
   		The next-to-last column shows the availability of UVCS observation for the event.}
   \label{cdaw_html}
\end{figure}

\begin{figure}
   \centering
   \includegraphics[width=10cm]{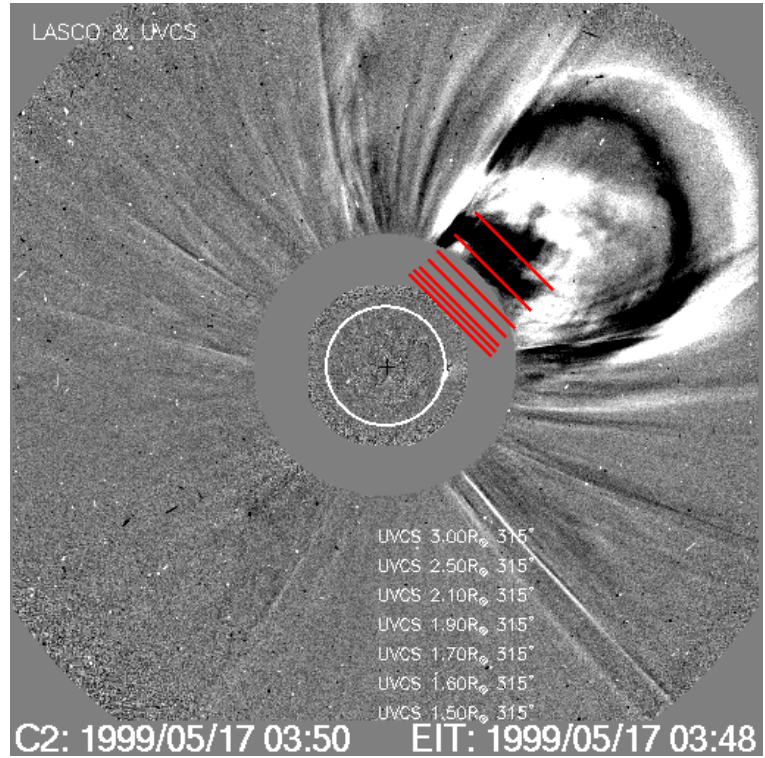}
   \vspace{0.75cm}
   \caption{An example of an image from the LASCO C2 running difference movie with the UVCS slits superimposed, 
   	these movies are included in the UVCS CME catalog web page of each event.}
   \label{lasco_uvcs}
\end{figure}

\begin{figure}
   \centering
	\includegraphics[width=5.5cm]{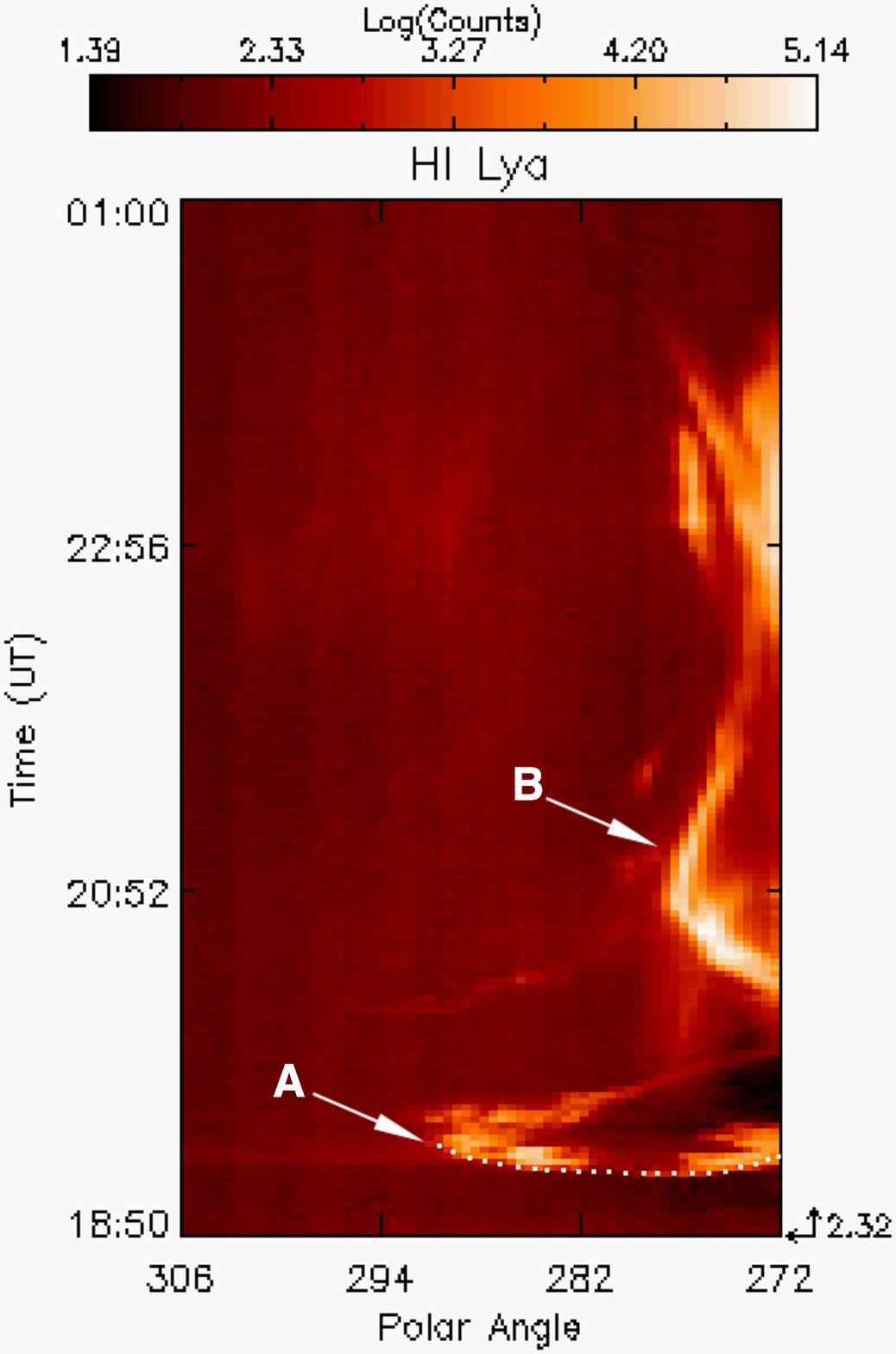} 
	\includegraphics[width=5.5cm]{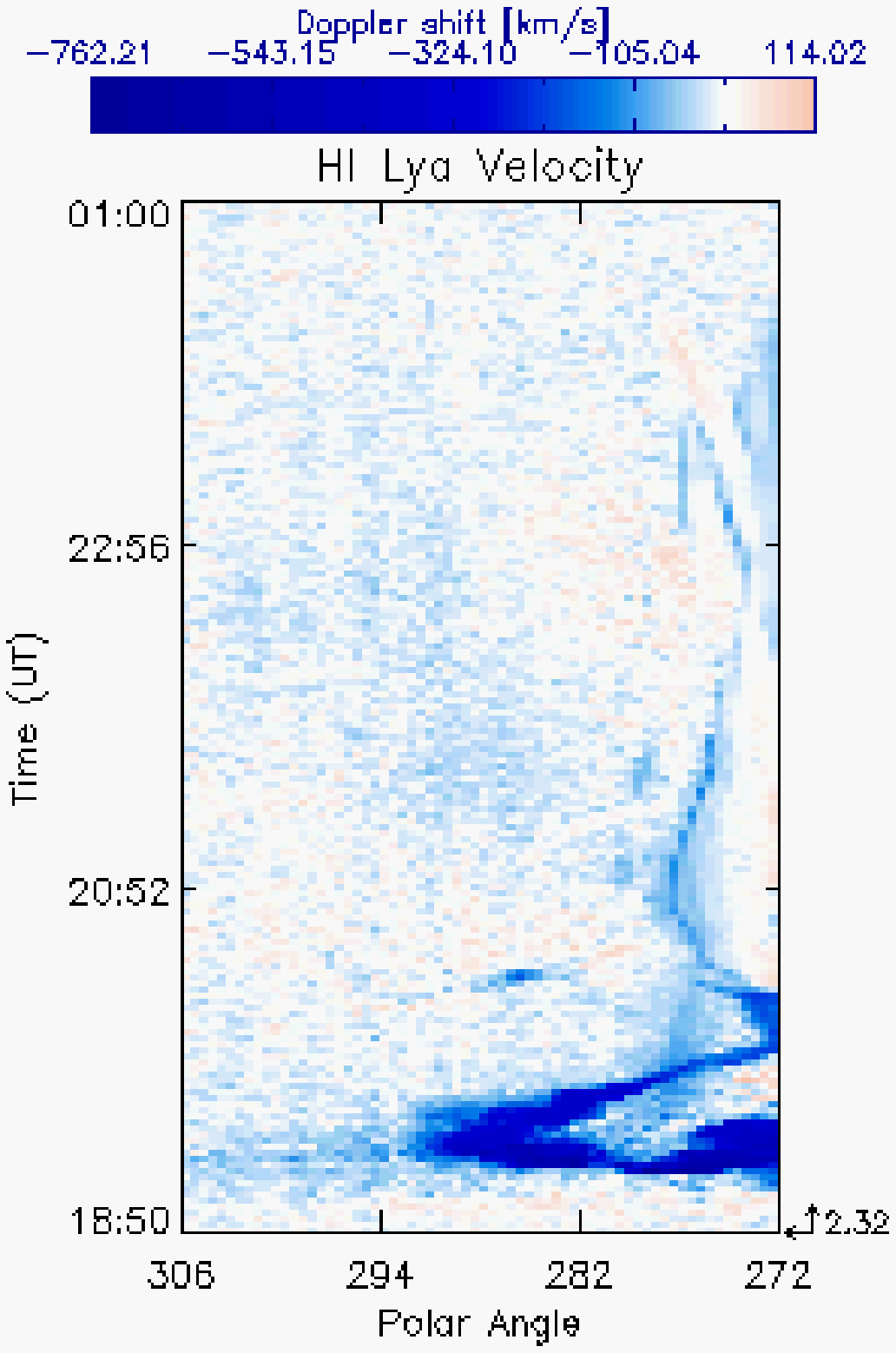} 
	\includegraphics[width=5.5cm]{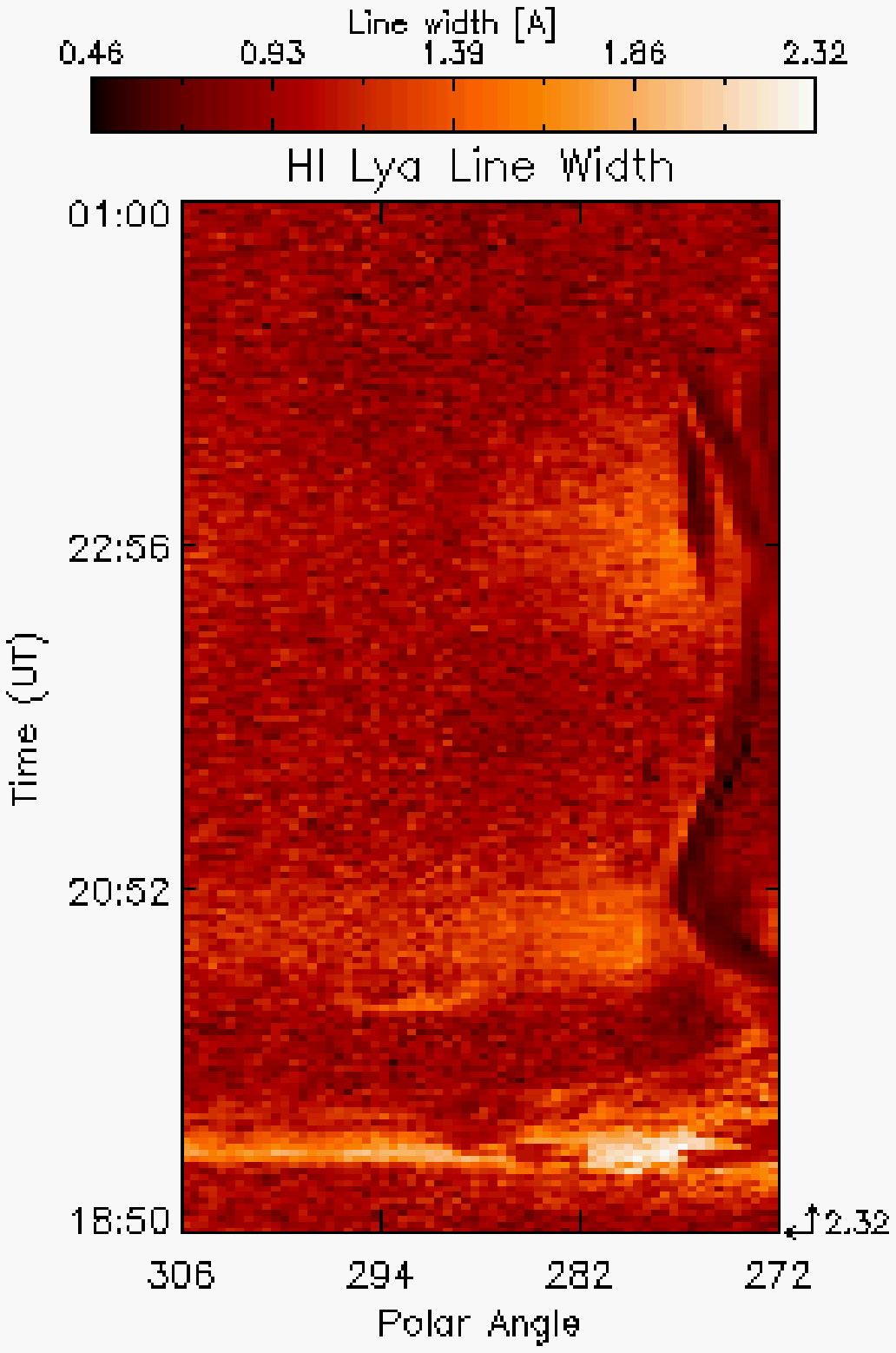}
  \caption{H~I Lya 1216\AA\/ intensity image (left panel), Doppler shift image (middle panel) and Line width image (right panel) for the 2000 June 28 event.
  Labels A and B in the left panel indicate the CME front and prominence/core material, respectively.}
  \label{cat_20000628}
\end{figure}

\begin{figure}
   \centering
	\includegraphics[width=5.5cm]{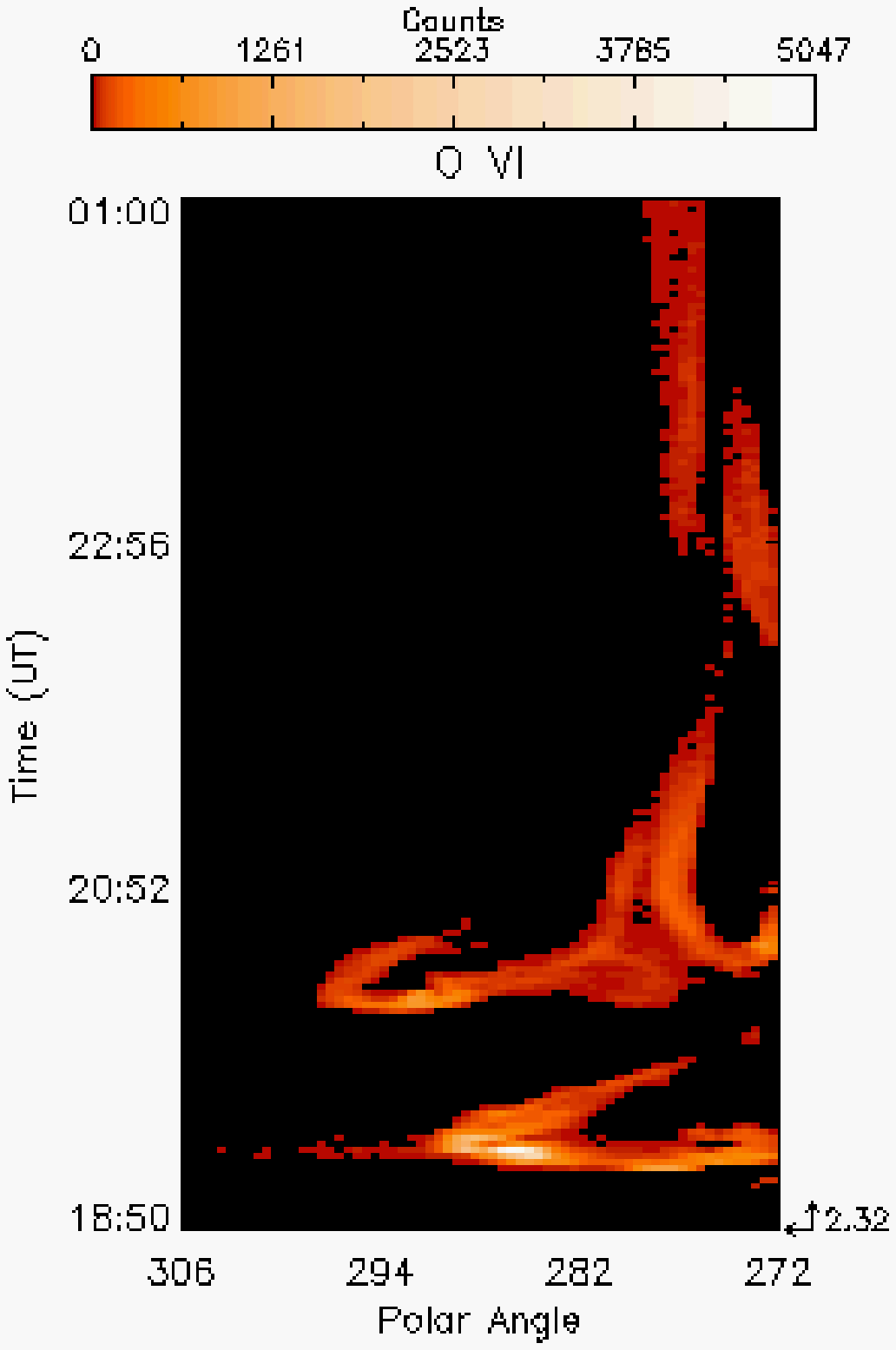} 
	\includegraphics[width=5.5cm]{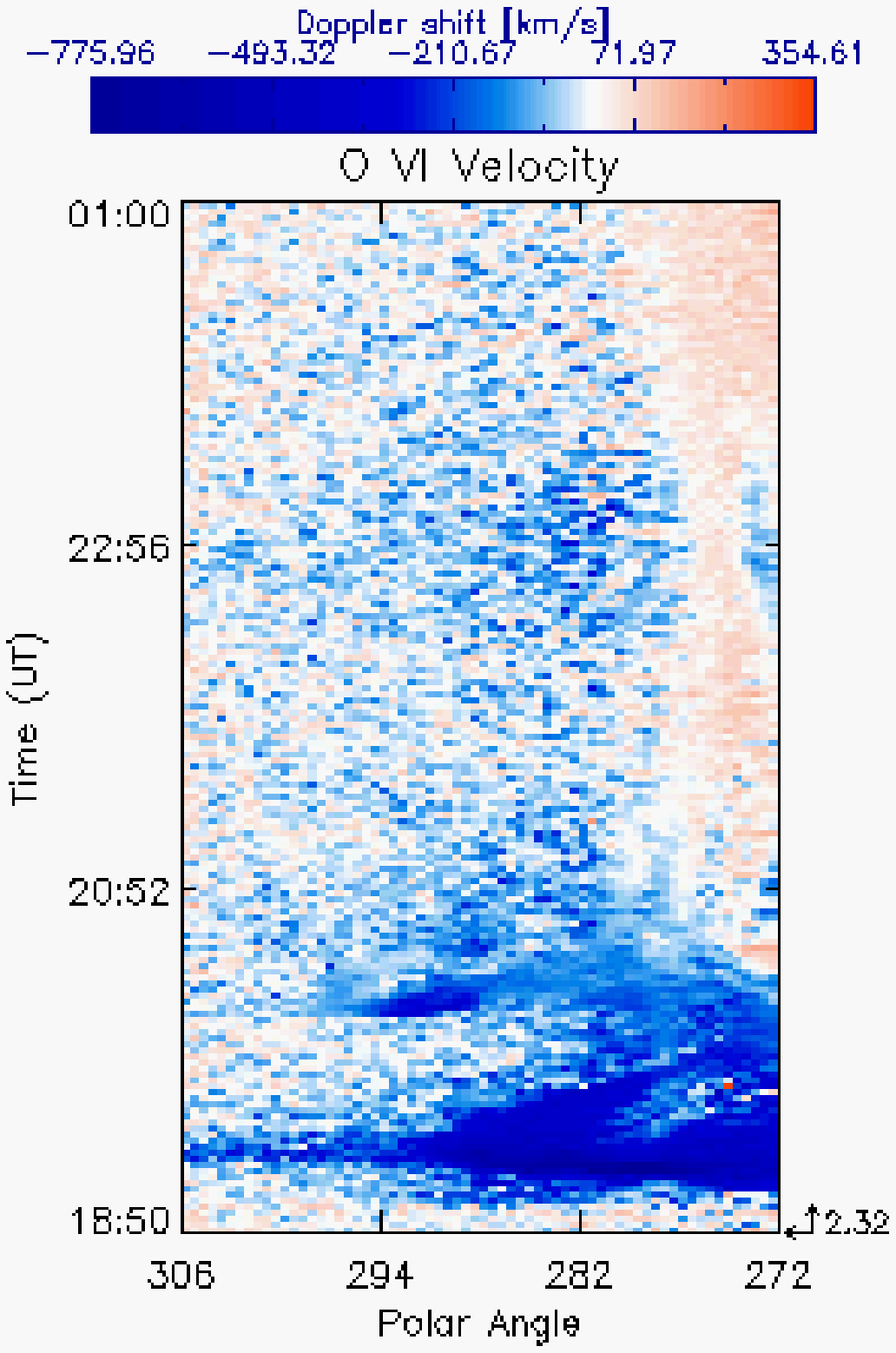} 
  \caption{O~VI 1032\AA\/ intensity image (left panel), Doppler shift image (right panel) for the 2000 June 28 event.}
  \label{cat_20000628_o6}
\end{figure}

The catalog of CMEs observed by UVCS is now available to the scientific community at the URL:
$\rm http://solarweb.oato.inaf.it/UVCS\_CME$. 
The top panel of Figure~\ref{cdaw_html} shows the main page of the catalog as a matrix of year and month of observation.
By clicking on a month in that matrix a list of the events is displayed, then by clicking on a single date the user opens a page  
containing some general information on the CME and a tutorial on the UVCS instrument and spectral analysis.  
It lists the UVCS data files, 
the time of first detection by UVCS, 
the heights at which the CME was observed, 
the maximum blue and red shift in km~s$^{-1}$,  
the wavelength ranges covered in the UVCS spectra and 
the lines in which the CME is detected. 
Each event page contains a frame from a movie of LASCO C2 difference images with the UVCS slit superimposed is also shown, 
see Figure~\ref{lasco_uvcs}.
A set of 2D images of the CME in selected lines give the line intensity during the observation as a function of time and the 
polar angle position along the slit measured counterclockwise from north pole.
Doppler velocity images can also be included, and they give the Doppler shift from the variation of centroid for selected spectral lines
with respect to the background corona as a function of the time and position along the slit.
Figure~\ref{cat_20000628} shows the 2000 June 28 CME \citep{Cia05} H~I Ly$\alpha$ 1216\AA\/ spectral line intensity image (left panel), 
the corresponding Doppler shift measured in H~I Ly$\alpha$ (middle panel) and the line width (right panel).  
The very bright, filamentary H~I Ly$\alpha$ emission is characteristic of ejected prominence material.  
The large line width early in the event results from bulk motion of the plasma along the line of sight, while the small line widths 
late in the event indicate proton temperatures far below coronal values.  
The velocity shown in the figure is computed from the line centroid with respect to the pre-CME corona, and the highest velocity material 
moves at -1800 km~s$^{-1}$ as inferred from the pumping of OVI doublet lines mentioned in Section~\ref{intro}.
In general the Doppler velocity maps from line centroid represent a lower limit of the line of sight velocity determined by computing
the maximum Doppler shift of the observed feature with respect
to 
the central wavelength value of the pre-CME spectral line emission. 
The same event is shown in the O~VI 1032\AA\/ line in Figure~\ref{cat_20000628_o6}.  
The overall structure is quite similar to that in H~I Ly$\alpha$, except that the bright O~VI feature at about 19:40 UT is nearly absent in H~I Ly$\alpha$.

\begin{figure}
   \centering
   \includegraphics[width=5.5cm]{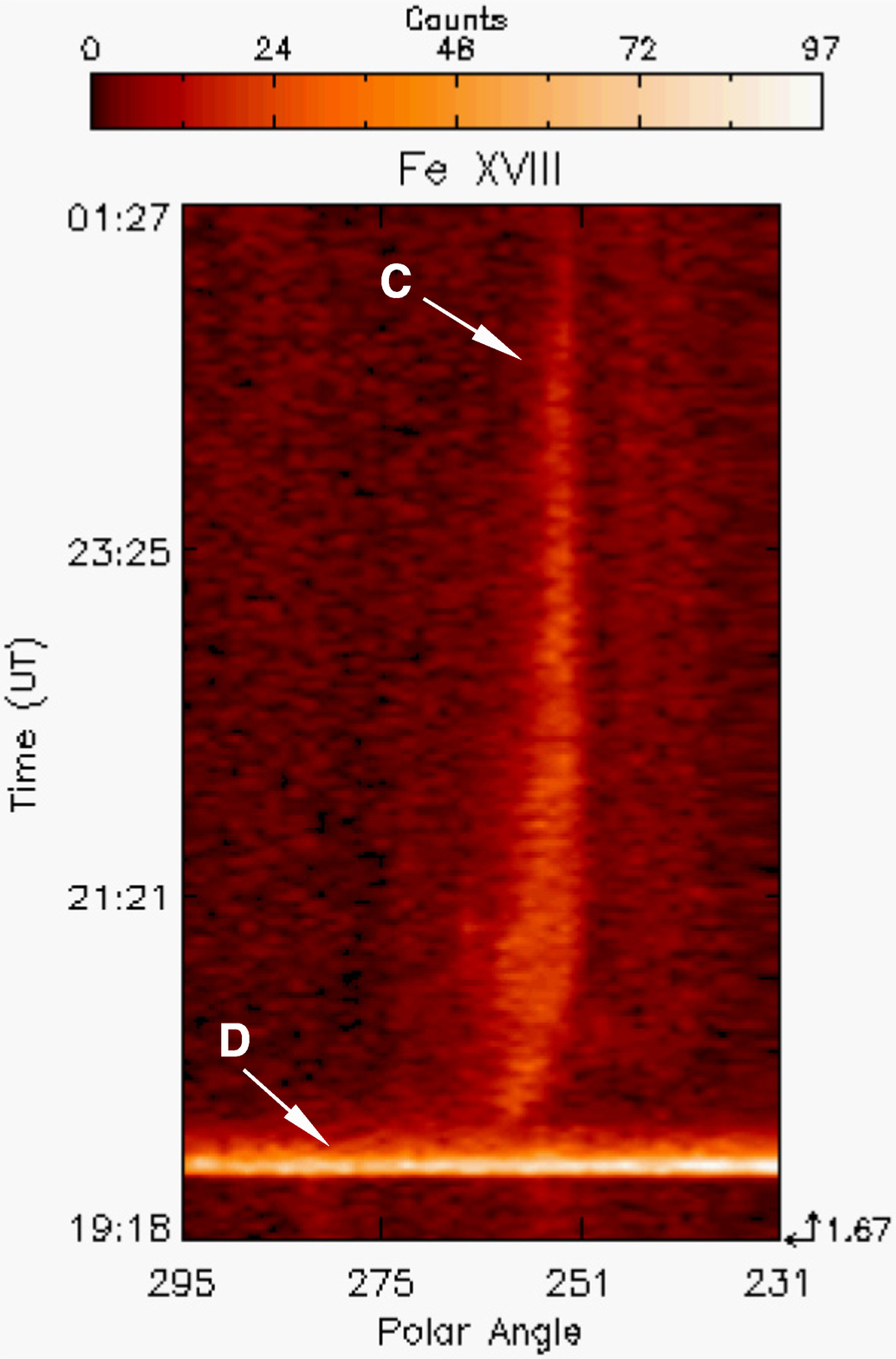}
   \includegraphics[width=5.5cm]{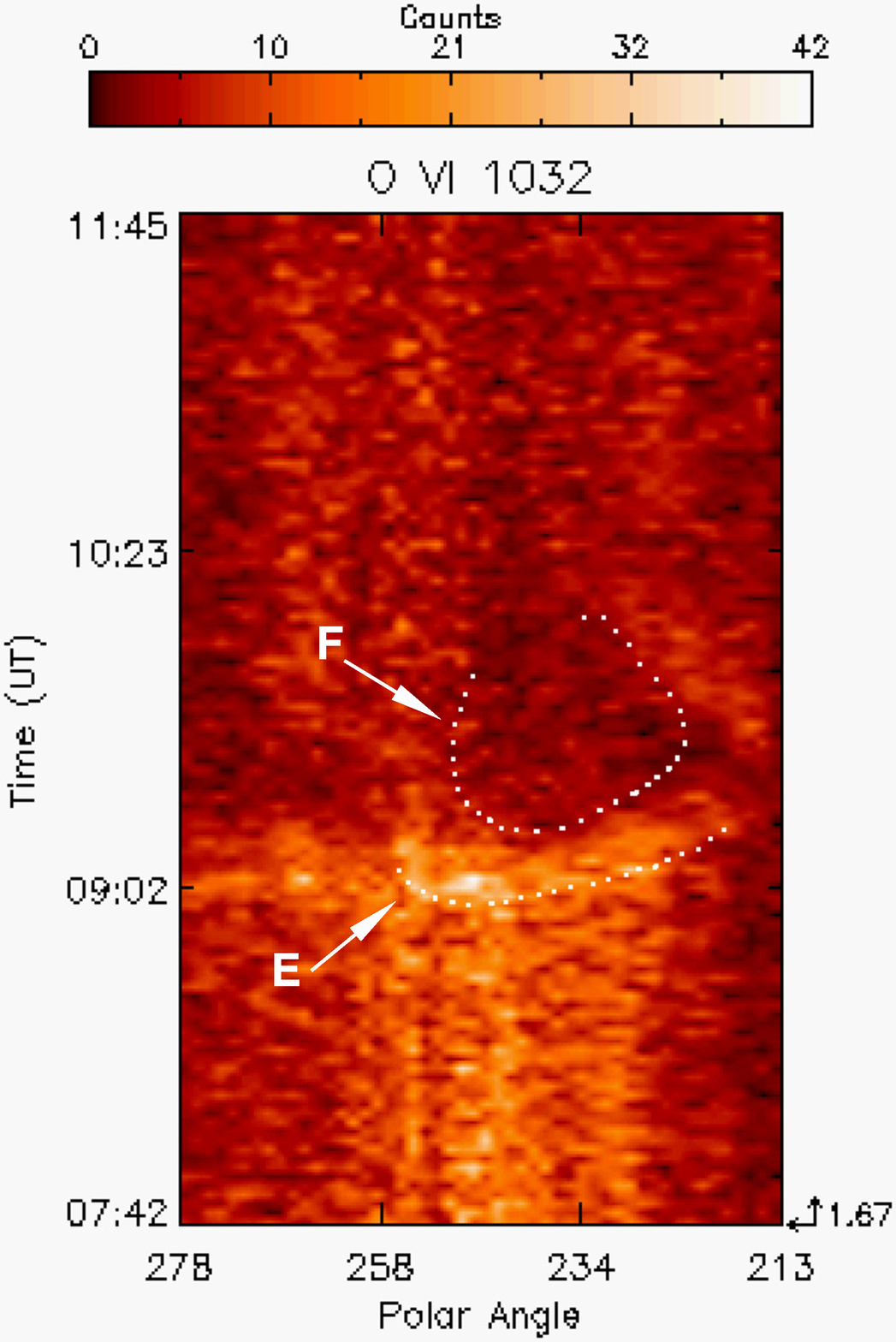}
   \includegraphics[width=5.5cm]{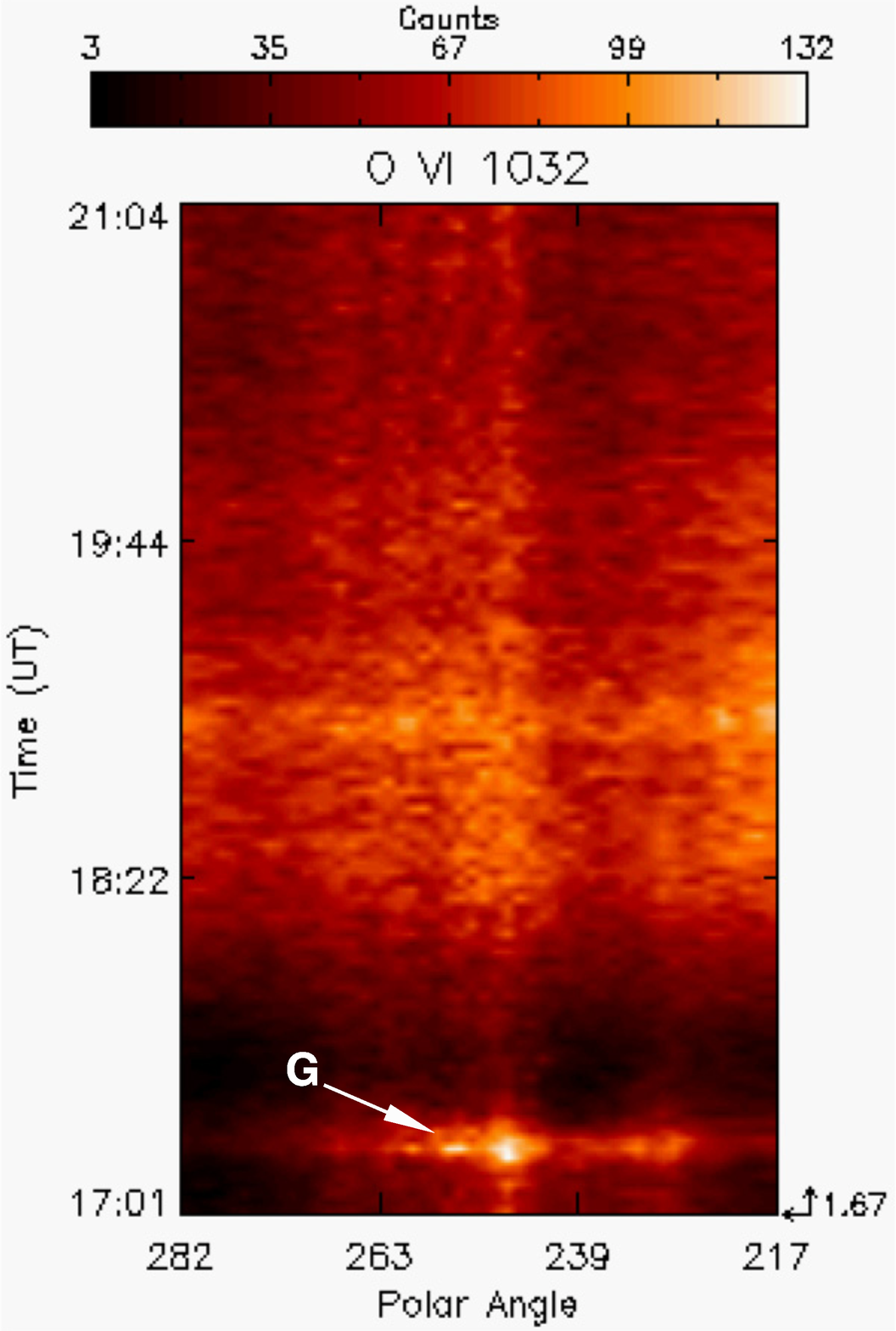}
   \caption{Left panel: Fe~XVIII 975\AA\/ intensity image of the current sheet 
   (C)
   on 2003 November~4. 
   The bright feature that fills the slit
   (D)
   is due to high background produced by X-rays from the powerful flare.
   Middle panel: O~VI 1032\AA\/ difference image for the 2003 November 2 CME detected by UVCS at 09:07UT.
   Front
   (E)
   and void
   (F)
   can be seen in O~VI 1032\AA\/ line.
   Right panel: O~VI 1032\AA\/ intensity image for the 2003 November 2 CME detected by UVCS at 17:12UT.
   Flare O~VI photons
   (G)
   scattered from coronal ions at 1.67R$_{\odot}$ are detected by UVCS from 17:12 to 17:29UT.}
   \label{uvcs_cs_diff_ovi}
\end{figure}

Left panel of Figure~\ref{uvcs_cs_diff_ovi} shows an image of a CS in the high temperature line of Fe~XVIII observed on 2003 November 4 \citep{Cia08}. 
The CS brightens at the time of the flare, moves southward by about $4^{\circ}$, narrows and fades over the course of many 
hours.  The feature that fills the slit at 19:48 UT is due to high background produced by X-rays from this exceptionally powerful flare.  
For the less intense events, difference images enhance the brightness of the CME material.  
An example is shown in the middle panel of Figure~\ref{uvcs_cs_diff_ovi} where the front and the void of the 2003 November 2 CME 
can be seen in O~VI 1032\AA\/ line. The O~VI emission brightens by about 20\% at 09:00 UT.
An example of the UV emission detected by UVCS from flare O~VI photons scattered from coronal O~VI ions is shown in the 
right panel of  Figure~\ref{uvcs_cs_diff_ovi}, the O~VI 1032\AA\/ line brightening from 17:12 to 17:29UT at 1.67~R$_{\odot}$ is associated to
a X class flare detected by GOES at the solar west limb \citep{Ray07}.  

The interpretation section of each event in the catalog draws a preliminary interpretation of the UVCS observations listing whether the 
front, void, shock, current sheet, prominence, flare O~VI, leg or helix were detected in the spectra (see section~\ref{interpretation}).
Finally the list of publications related to the event and some comments are included in each single event page. 

All information contained in the online UVCS CME catalog is also included in the CDAW LASCO CME catalog, 
available at the URL: $\rm http://cdaw.gsfc.nasa.gov/CME\_list$. 
In the page with the monthly list of LASCO CMEs, shown in the bottom panel of Figure~\ref{cdaw_html}, 
UVCS is listed in the next to last column for the events detected. 
By clicking on UVCS the user will get the same page as in the UVCS CME catalog, described so far.

The online catalog can also be searched to select events in a user--given range of dates, velocities, latitudes and Doppler shifts,
and/or to select events probably connected to X flares and type II radio burst, 
or interpreted as signature of, for example, shock, current sheet, O~VI flare etc.,
the catalog returns the list of the only CMEs meeting the search criteria in a table similar to the monthly list in the catalog.

\subsection{CME Statistics}

The total number of CMEs detected by UVCS from 1996 April 1 to 2005 December 31 is 1059, including 
16 events, reported in Table~\ref{uvcsonly},  detected by UVCS only and therefore not listed in the LASCO CME catalog.
Over these events, 5 are so called narrow CME not definitively place into jet-like or CME classification \citep{Dob03}. 
For 4 other events, reported in Table 1, there is not a clear evidence of relation to CME in LASCO catalog. 
In the same time period the LASCO catalog used as basis of this work contains 10491 CMEs.
It is important to note when comparing UVCS and LASCO 
CME
observations that UVCS observed mainly collisionally excited lines, whose intensity is proportional to 
$n_e^2$, 
while LASCO detects light from the solar disk that is scattered by electrons, so that the intensity is proportional to 
$n_e$.
For this reason massive fronts or dense prominence material  can produce a larger increase of the UV compared with 
the visible emissivity. Otherwise, in case of faint events the increase of collisional emission should be balanced 
by the strong dimming of the radiative one, reducing the chance of detecting the CME in the UV.

\begin{figure}
	\centering
	\includegraphics[angle=90,width=8.5cm]{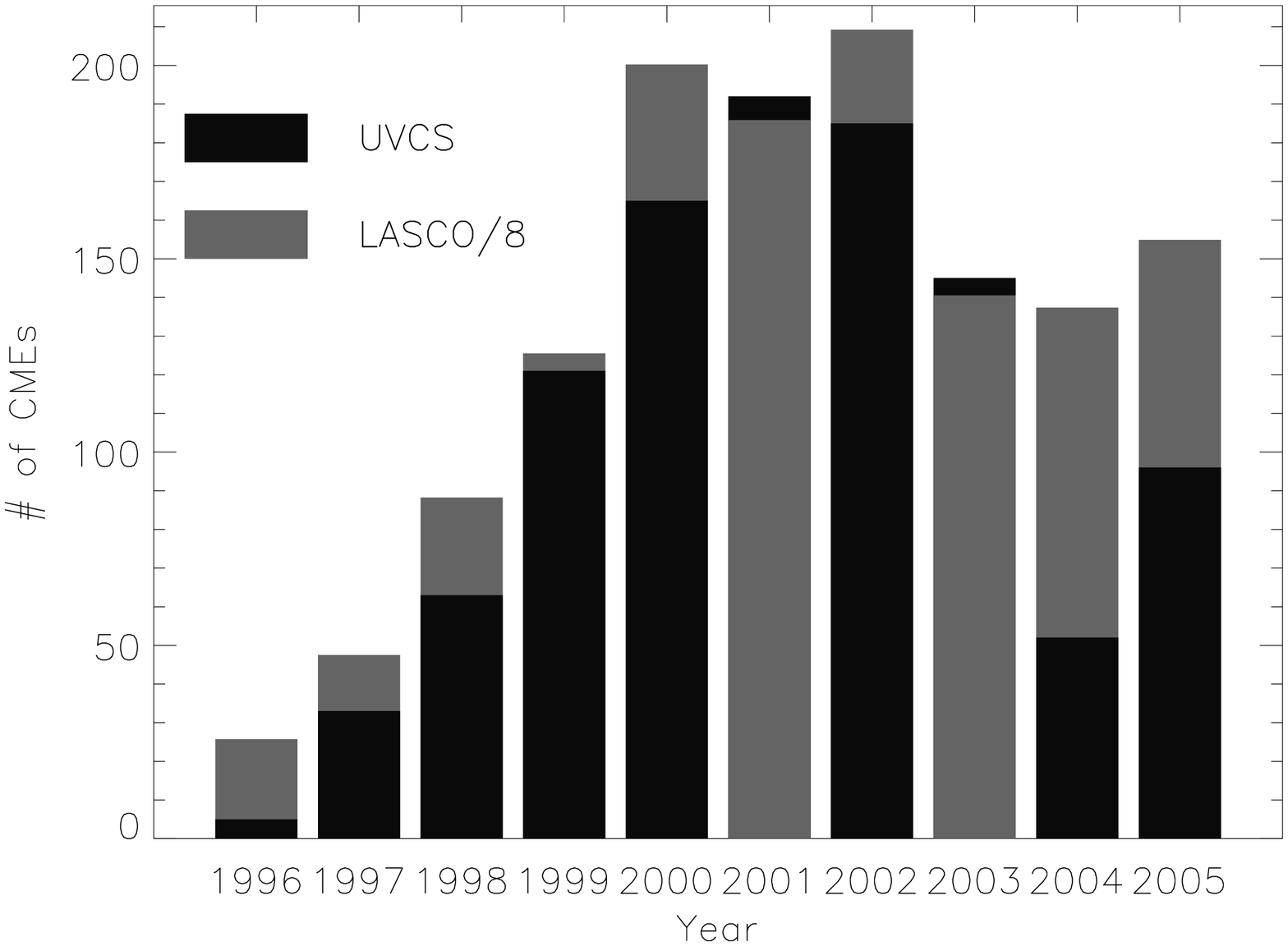} 
	\includegraphics[angle=90, width=8.5cm]{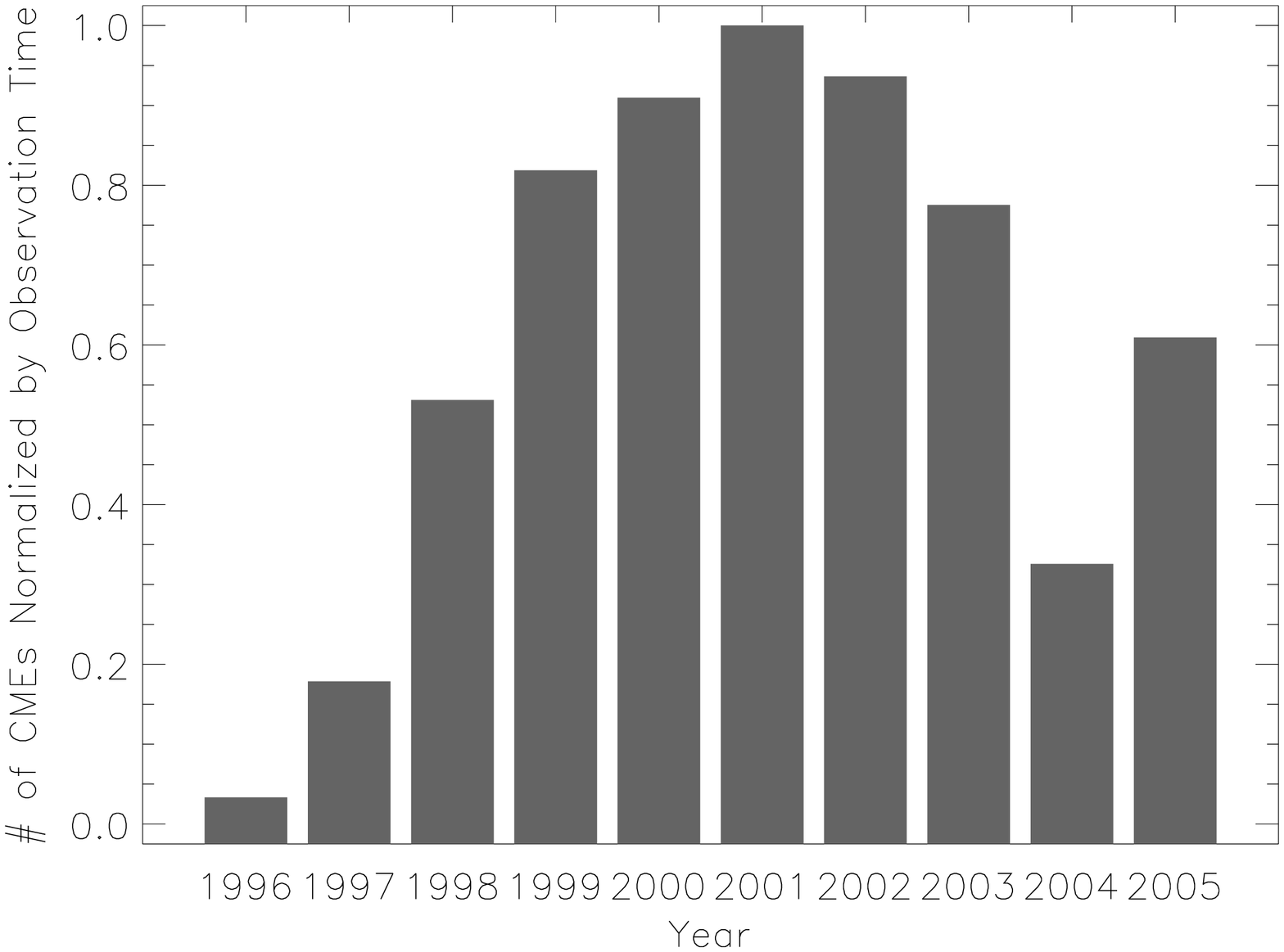}
  \caption{Number of CMEs per calendar year observed by UVCS (black) and LASCO (grey) 
  		during the period 1996 - 2005 (left panel). The number of LASCO CMEs is divided by 8.
           	The right panel shows the normalized number of CMEs divided by the total observation time per year.}
  \label{cme_yrn}
\end{figure}

The distribution of the events as a function of the calendar year is shown in Figure~\ref{cme_yrn}. 
The left panel of the figure shows the total number of CMEs per year as compared to the LASCO CMEs divided by 8, 
and the right panel the normalized number of CMEs divided by the total observation time per year (given in the top panel 
of Figure~\ref{obs_summary}).
While the distribution of UVCS CME detections as a function of the solar cycle is very similar to LASCO, UVCS
detected about one tenth of the LASCO CMEs. The UVCS field of view given by the entrance slit, typically spans 
about $1/8$ of the 360$^\circ$\/ range covered by LASCO and the duty-time of UVCS is lower than LASCO because of the 
pointing and grating mechanism movements.
Nevertheless, there are 
many hundreds of
events quite bright in visible light not detected in the UV emission despite the 
apparently coincident
UVCS pointing;
indeed a detailed analysis of these cases, out of the scope of this paper,
is needed in order to understand if the combined effect of CME speed and density makes no significant changes on the spectra 
or very small changes that were missed in our search.
Moreover, some CMEs detected by LASCO are very faint and UVCS observes preferentially CME with higher mass and energy 
as shown in Section~\ref{mass}. 

In the first year of scientific operation UVCS observed for about 9 months during the minimum of solar activity
catching
only 5 CMEs, later the annual CME rate increases from about 30 CMEs in 1997, still at solar minimum, to about 190 during
solar maximum, in 2001 and 2002, then the rate slowly decreases in the declining phase of solar cycle.
The lack of events in 2004 is probably due to the large time spent observing at high latitudes, where CMEs are 
detected by UVCS and LASCO less frequently
(see below).

\begin{figure}
   \centering
   \includegraphics[angle=90, width=8.5cm]{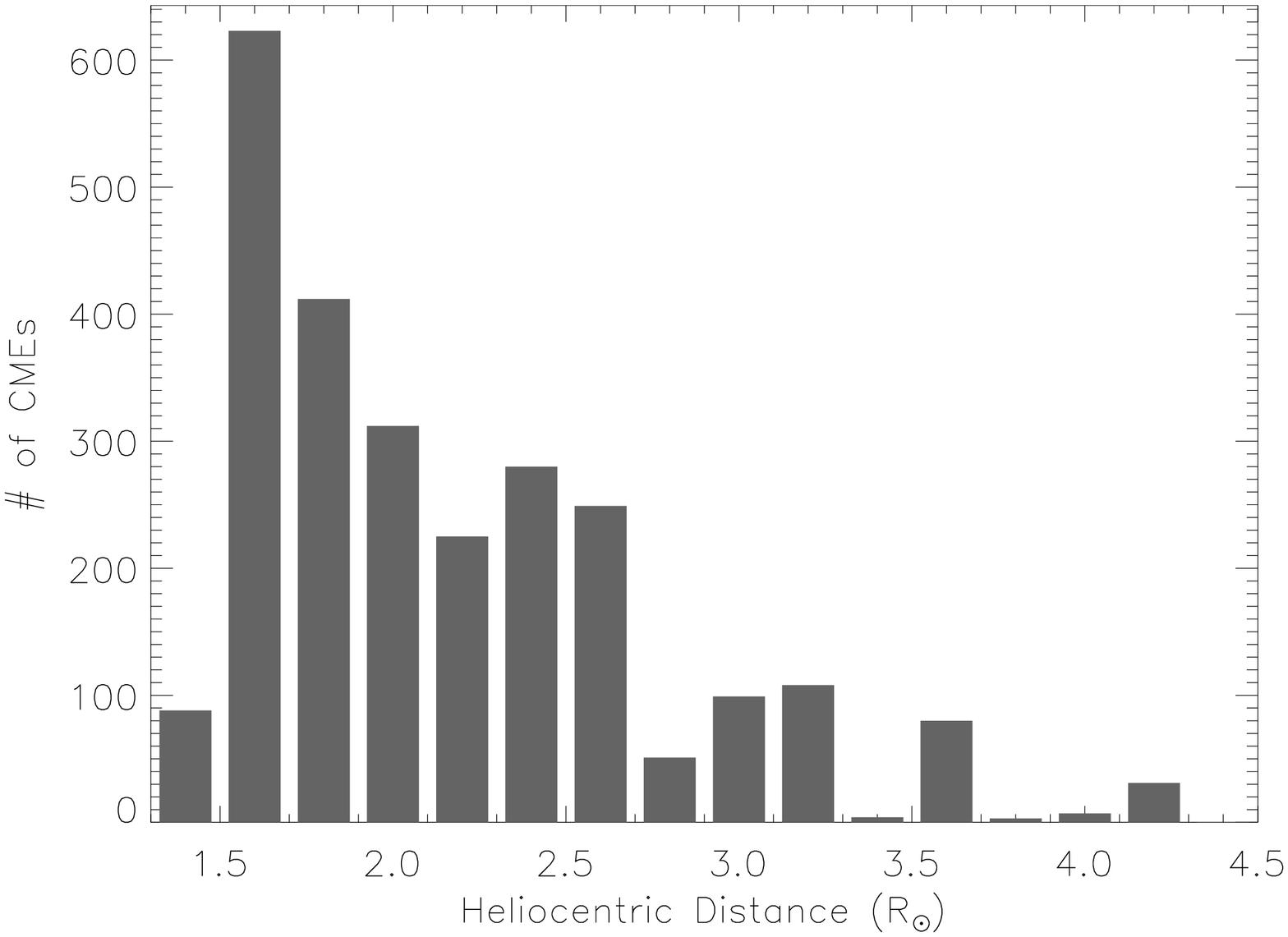} 
   \includegraphics[angle=90, width=8.5cm]{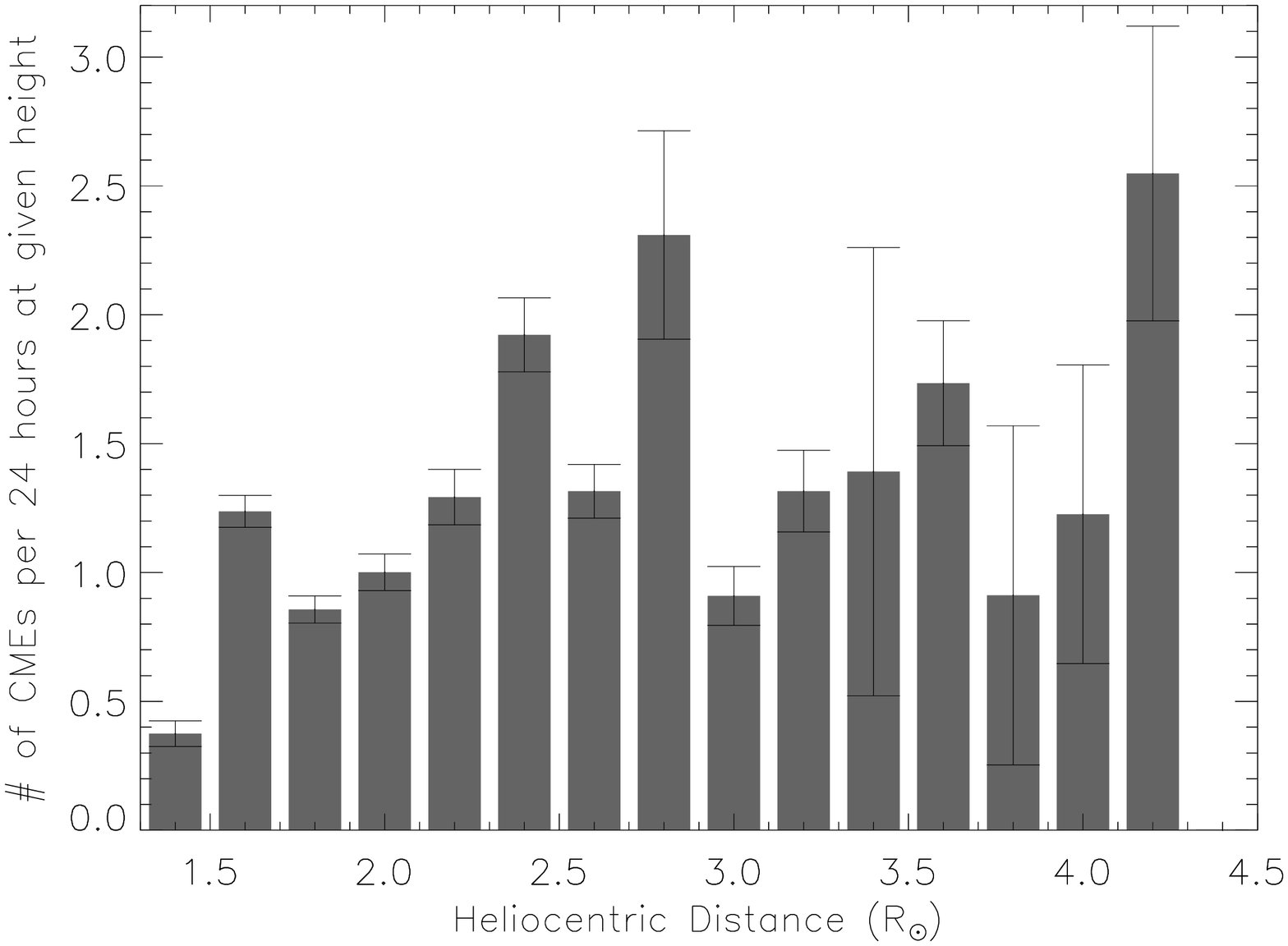}
   \caption{Number of detections of CME material as a function of the heliocentric heights (left panel). 
    The number of CMEs observed by UVCS per day at each height (right panel). 
    The height range is binned by 0.2~R$_{\odot}$. 
    The heights with large uncertainty are those less used for UVCS observations.}
   \label{cme_h}
\end{figure}

The instantaneous field of view of UVCS covers a 40$^\prime$ strip at a given polar angle and heliocentric height,
as expected, most of the CMEs were detected at the heights where UVCS spent the most observing time, i.e. 1.6 to 1.7~R$_{\odot}$\ 
(see also bottom panel of Figure~\ref{obs_summary}).
The number of detections as a function of height, binned by 0.2~R$_{\odot}$\ is shown in the left panel of Figure~\ref{cme_h}. 
Note that since the same CME can be detected at several heights, the total number of detections is larger than the total number of CMEs.  
Each CME on average is detected at 2 or 3 different heights, and in some cases (e.g. \citet{Cia00, Bem07}) it has been possible to determine 
for a single event the evolution with the heliocentric distance of physical parameters determined from UV diagnostics.  

There are 31 CMEs detected at heights larger than 5.0~R$_{\odot}$, and the largest height at which a CME was detected is 
7.9~R$_{\odot}$\ (1997 August 13).  At large heights, the slit covers a smaller angular fraction of the field of view, and CMEs are 
much fainter because of the rapid falloff of density as the plasma expands.
This is partly compensated by an increase in the sensitivity of UVCS with height \citep{kohl95}.
In addition, UVCS observations at large heights are usually made
keeping the same height for long time, collecting a large number of exposures, that makes faint events more easily detected.
On the other hand, the observations below 1.5~R$_{\odot}$\, usually consist of 
a
few short exposures enough to obtain 
good statistics,
but not optimized for CME detection, 
and
for this reason we have a small number of CMEs around 1.4~R$_{\odot}$.
In the right panel of Figure~\ref{cme_h} we plot the number of CMEs detected by UVCS per observation days (24 hours) spent at each 
height (the observation time distribution as a function of height is given by the bottom panel of Figure~\ref{obs_summary}).
This plot gives the average expected number of CMEs detected by UVCS per observation day as a function of height. 
Thus, within the statistical uncertainties, the probability of catching a CME with UVCS does not depend on the observation height. 
 
\begin{figure}
   \centering
   \includegraphics[angle=90, width=10cm]{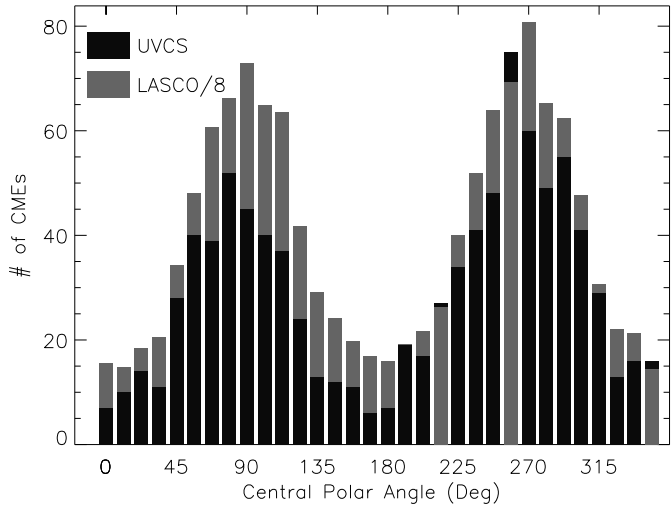}
   \caption{Distribution of the number of CME detected by UVCS (black) and LASCO (gray) as a function 
   		of the CME central polar angle. The number of LASCO events is divided by 8.}
   \label{cme_pa}
\end{figure}

The distribution of the number of UVCS CMEs as a function of the CME central PA is very similar to that obtained with 
LASCO.  In Figure~\ref{cme_pa} UVCS and LASCO distributions are compared (LASCO numbers are divided by 8).
The number of CMEs detected decreases at high latitudes, although about 6\% of the events are located 
within $\pm$15$^{\circ}$ around the poles.

\begin{figure}
   \centering
   \includegraphics[angle=90, width=10cm]{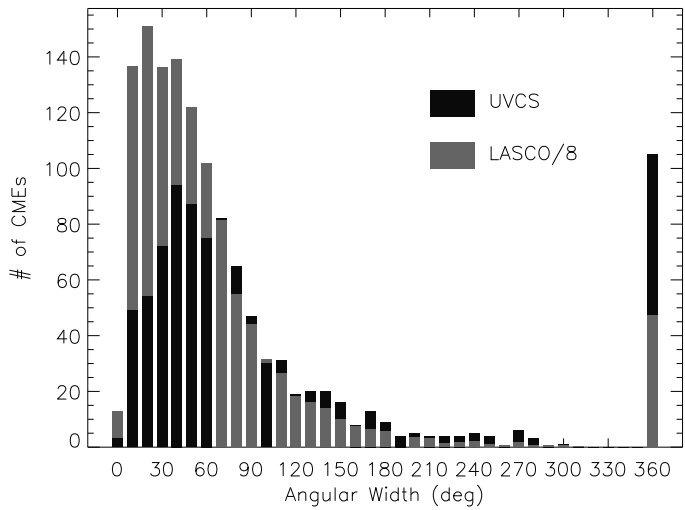}
   \caption{Distribution of the number of CME detected by UVCS (black) and LASCO (gray) as a function 
   		of the CME angular width. The number of LASCO events is divided by 8.}
   \label{cme_widt}
\end{figure}

Figure~\ref{cme_widt} shows the number of UVCS CMEs as a function of the angular width determined by LASCO. 
Excluding the halo events, the average angular width of LASCO CMEs is about 55$^{\circ}$, while it is about 75$^{\circ}$ 
for the sub-sample detected by UVCS. Counting CMEs with width $\le$120$^{\circ}$, as \citet{Gop09}, because the true
angular size is better defined for narrow events, the average apparent width is about 45$^{\circ}$ for LASCO and 
about 56$^{\circ}$ for UVCS sub-sample.
UVCS detects preferentially the CMEs with large angular size as they involve a larger portion of the solar corona and 
are therefore more likely to enter the UVCS field of view. Indeed, in general UVCS detects about 1/10 of LASCO CMEs, 
while in case of halo CMEs the ratio is 1/4. More precisely, the number of halo events detected by UVCS is 106,
which represents the 10\% of the sample, while only about 3\% of LASCO CMEs are classified as halo events. 
A study of the properties of halo CMEs observed by UVCS has been performed by \citet{Cia06} on a preliminary sample 
of 22 events.
On the other hand, events with small apparent angular width are alternatively listed as coronal jets or narrow CMEs 
\citep{Dob00, Dob03, Bem05}, because they could be triggered by different mechanisms, therefore not all of this kind of events
are included in the UVCS and LASCO CME catalogs.

\subsection{UVCS CME Dynamical Properties}

Each CME in the LASCO catalog is characterized by a linear speed obtained by fitting a straight line to the height-time measurements; 
the linear speed serves as an average speed in the plane of the sky within the LASCO field of view.

\begin{figure}
   \centering
   \includegraphics[angle=90, width=10cm]{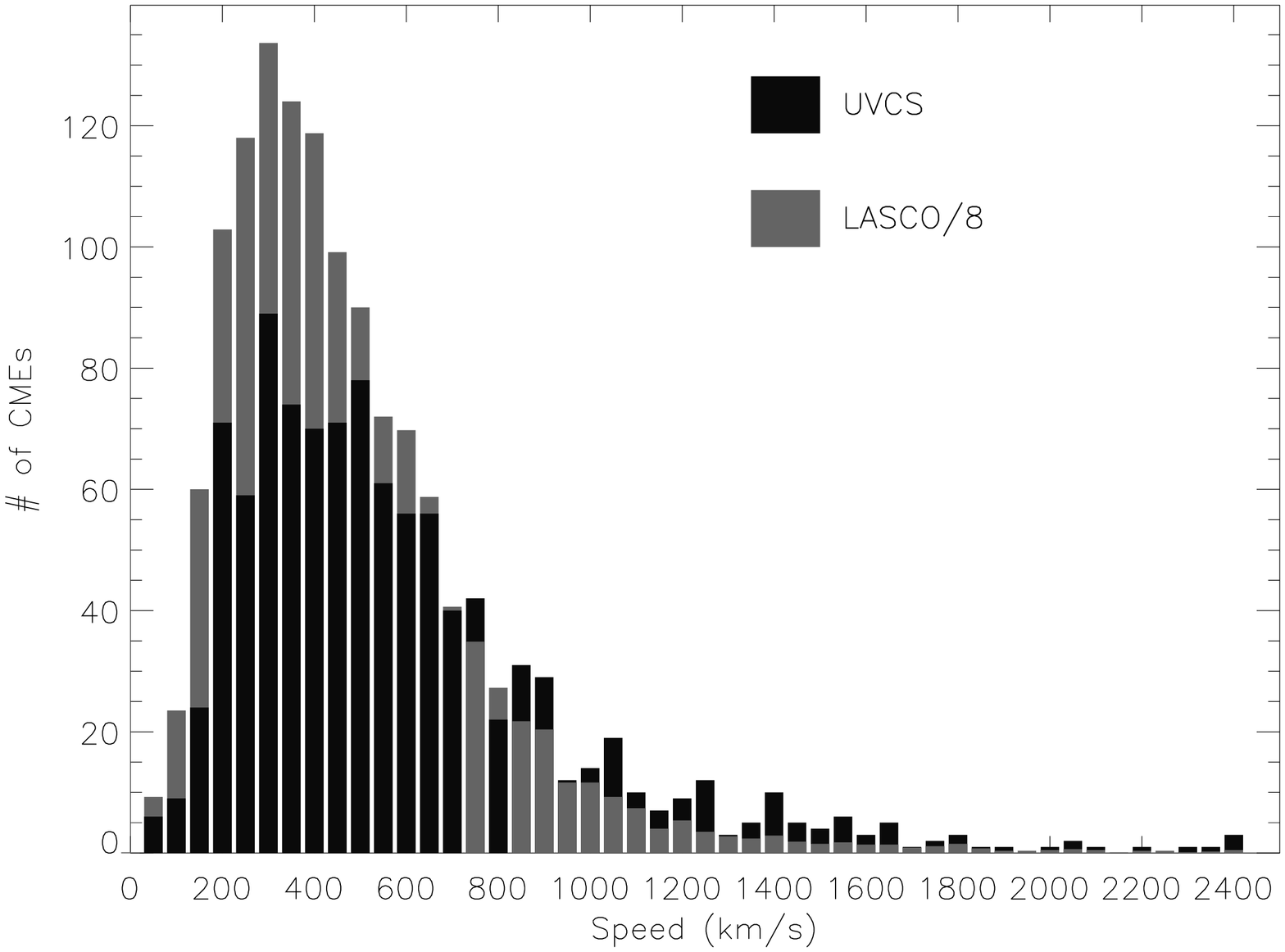}
   \caption{Number of CME observed by UVCS (black) and LASCO (gray) as a function of the CME linear speed.
		The number of LASCO events is divided by 8.}
   \label{cme_speed}
\end{figure}

The distribution of UVCS CMEs as a function of LASCO linear speed is plotted in Figure~\ref{cme_speed}.
In black are UVCS data and in grey LASCO data divided by 8.
The average 
linear
speed of the 
sub-sample detected by
UVCS 
is 596~km~s$^{-1}$  while for 
the whole
LASCO
catalog
it is 472~km~s$^{-1}$.
This may be related to the fact that high speed CMEs tend to be wider \citep{Yas04} and therefore more probable 
to be detected by UVCS.  
Moreover, UVCS detects preferentially more massive CMEs (see Section 3.3) which we tend to be 
slightly faster, as we verified by studying the CME speed distribution as a function of CME mass over the whole LASCO catalog.

\begin{figure}
   \centering
   \includegraphics[angle=0, width=8.5cm]{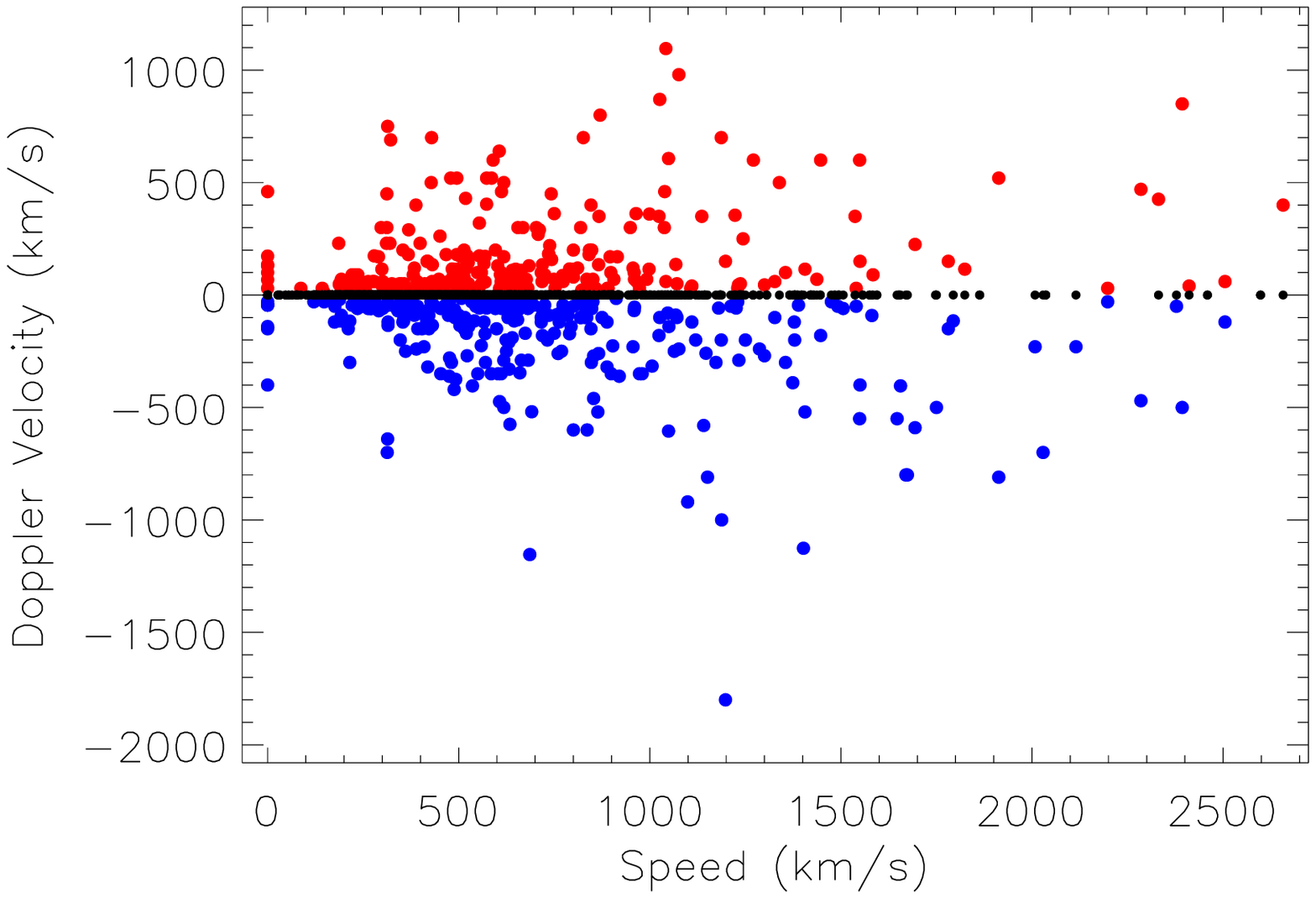} 
   \includegraphics[angle=0, width=8.5cm]{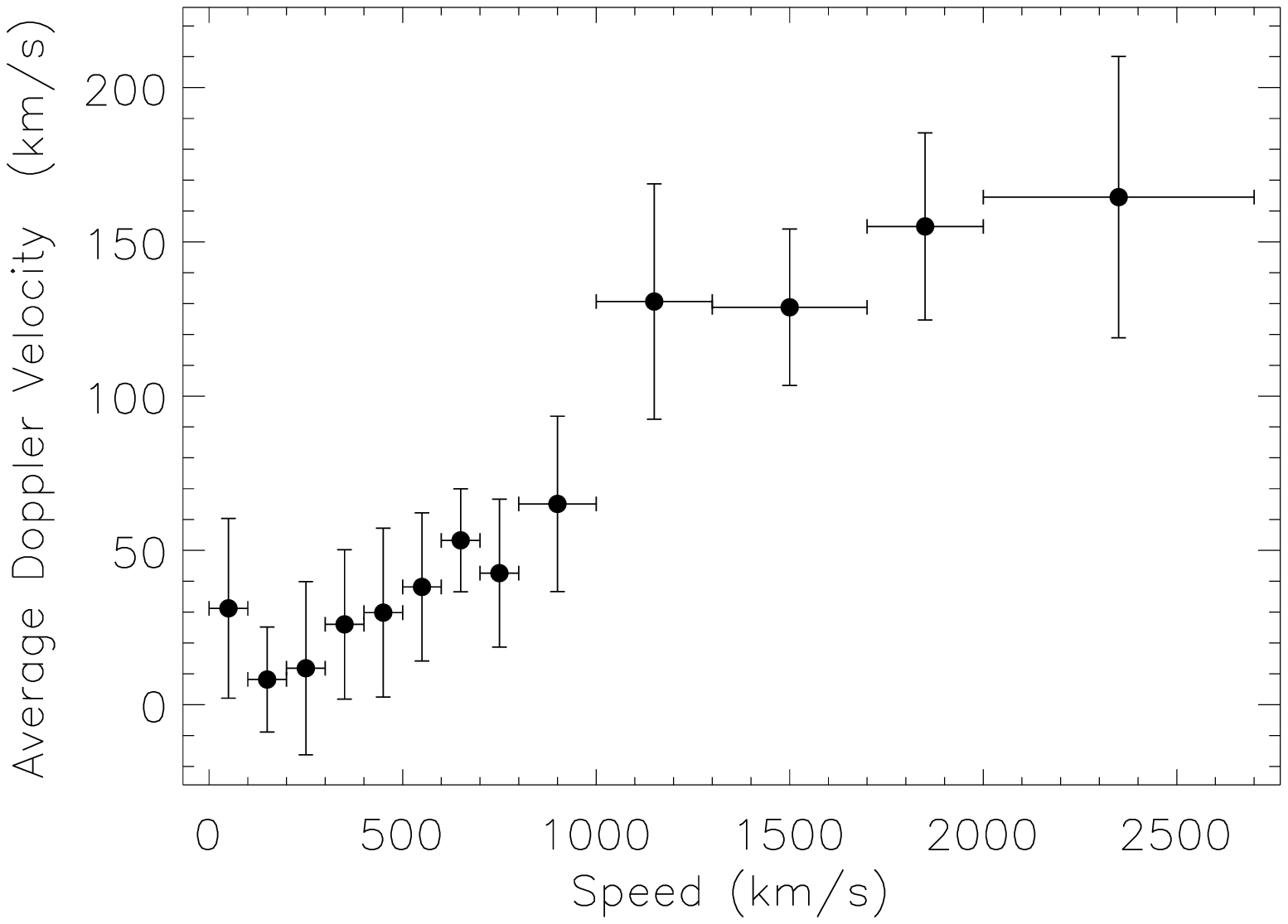}
\caption{Left panel: line of sight velocity as obtained from Doppler shifts for each event as a function of the CME speed in the plane of the sky 
   		determined by LASCO. 
		Blue and red bullets show the maximum blue (incoming speed) and red shifts detected for each observed event respectively. 
		Black bullets correspond to events for which the centroids of the detected spectral line do not show evident shift from pre-CME value. 
		The absolute value of the Doppler velocity averaged over LASCO C2 linear speed is in the right panel. 
		The bin size increases with LASCO speed, 
		as shown by the x-axis uncertainties.}
   \label{cme_dopp}
\end{figure}

From the analysis of the variation of the spectral line centroids at the time of CMEs crossing the UVCS slit it has been possible 
to determine the Doppler shifts, i.e. the line of sight velocity, for about 40\% of the events detected by UVCS.
The maximum Doppler velocity detected for each event is plotted in the left panel of Figure~\ref{cme_dopp} as a function of the LASCO 
linear speed, which represent the CME velocity in the plane of the sky. 
A wide range of Doppler velocities are detected at all LASCO linear speeds. 
Over the 1059 events detected by UVCS, 262 CMEs show red shift (speed away from the Earth), 
268  blue shift (speed toward the Earth) and 106 events show both red and blue shift.  
Although the left panel of Figure~\ref{cme_dopp} shows a wide range of Doppler shifts for each LASCO speed, the larger line of 
sight speeds correspond to the larger linear speeds as shown in the right panel of Figure~\ref{cme_dopp}.
In this plot we show the absolute values of blue and red Doppler velocity averaged over each LASCO speed bin. 
The bin size increases with LASCO speed in order to increase statistical significance at larger speeds ($\ge$1000~km~s$^{-1}$), 
where only few events are detected.  At first glance, the small ratio of Doppler shift velocity to LASCO speed suggests motions very close
to the plane of the sky.  However, the LASCO speed pertains to leading edge, while most of the UVCS Doppler shifts
pertain to the slower prominence material in the CME interior.

\subsection{UVCS CME Mass and Energy}\label{mass}

LASCO observations detect the photospheric light scattered by the CME electrons and provide a means to measure plasma density, 
therefore, with a number of assumptions a representative value of the CME mass \citep{Vou10}.
The kinetic energy then is obtained from the linear speed and the mass.
There are generally large uncertainties in these values, for example, \citet{Vou10} estimated that CME masses may be underestimated 
by a factor of 2 and CME kinetic energies by a factor of 8. In this paper, concerning CME mass and energy, we are focusing 
on comparison of the distributions from the whole LASCO and UVCS sub-sample catalog.

\begin{figure}
   \centering
   \includegraphics[angle=90, width=10cm]{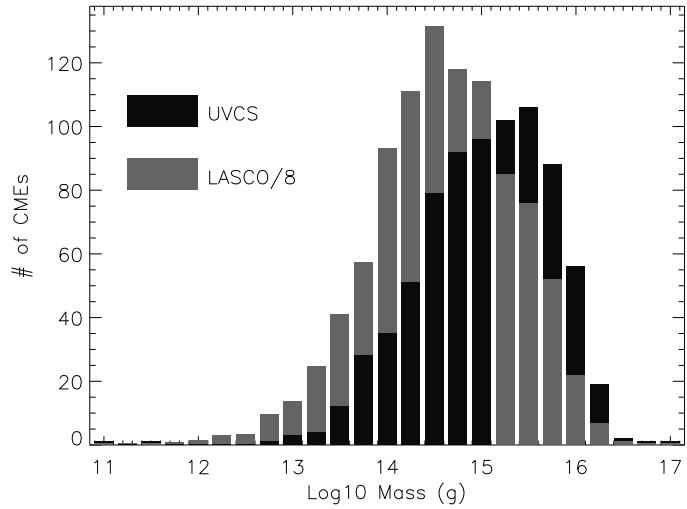}
   \caption{Number of CME observed by UVCS (black) and LASCO (gray) as a function of the CME mass.
		The number of LASCO events is divided by 8.}
   \label{cme_mass}
\end{figure}

Figure~\ref{cme_mass} shows the mass distribution of CMEs  determined by LASCO for non-halo events, 
during the period 1996--2005 (gray bars) and that for the events in the UVCS catalog (black bars).
The average mass of the events detected by LASCO is 3.89 $\times 10^{14}$ g while for the sub-sample of CMEs 
detected by UVCS it is 1.05 $\times 10^{15}$ g.  The distribution shows that UVCS detects more massive CMEs, as massive events are 
brighter and/or wider making them more easy to be detected by UVCS. 
High density CME cores are often detected at large heights as they remain bright in spite of their expansion. 
Since the LASCO catalog does not give mass estimates for halo events, and halo events are more likely than other CMEs to be seen 
with UVCS, the average mass of UVCS CMEs given above is probably an underestimate.
      
\begin{figure}
   \centering
   \includegraphics[angle=90, width=10cm]{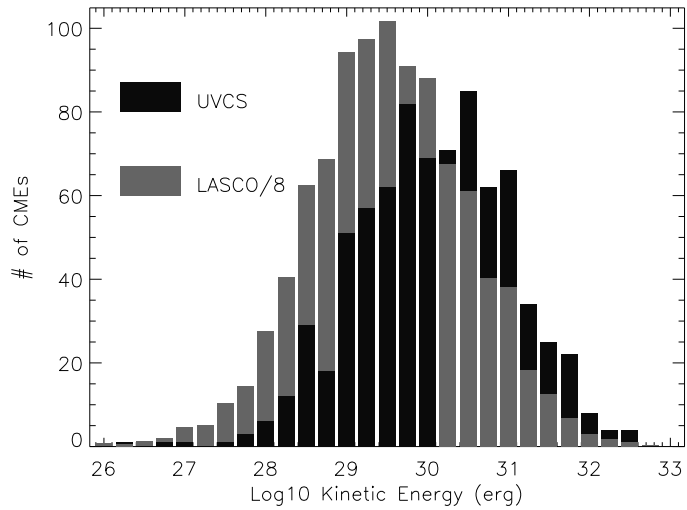}
   \caption{Number of CME observed by UVCS (black) and LASCO (gray) as a function of the CME kinetic energy.
		The number of LASCO events is divided by 8.}
   \label{cme_ke}
\end{figure}

The distribution of the UVCS CMEs with the kinetic energy computed from LASCO is shown in Figure~\ref{cme_ke}.
As a consequence of Figure~\ref{cme_speed} and Figure~\ref{cme_mass} the average kinetic energy of the UVCS CMEs 
(1.22$\times 10^{30}$ erg) is more than three time higher than that of LASCO CMEs (3.26$\times 10^{29}$~erg).   

\subsection{Spectral Lines and Intensities}

Table~\ref{tablines} lists the ions that have been detected in CMEs by UVCS, the peak of formation temperature and the 
wavelengths of their spectral lines \citep{der97,lan12}.
Some of the listed lines are observed much more frequently than others.  
The H~I Ly$\alpha$ line is detected in about 78\% of the events, the O~VI doublet in about 73\%, and H~I Ly$\beta$ and 
C~III 977\AA\, lines in about 20\%.
This depends on the physical properties of the ejected material such as temperature, density and outflow speed, 
but also on the selected UVCS wavelength range 
(i.e. spectral detector masks).
The latter almost always includes the O~VI doublet and H~I Ly$\alpha$ lines as the UVCS detectors were designed to have 
high sensitivity at those wavelengths.

\begin{figure}
\centering
\includegraphics[width=6.25cm,angle=90]{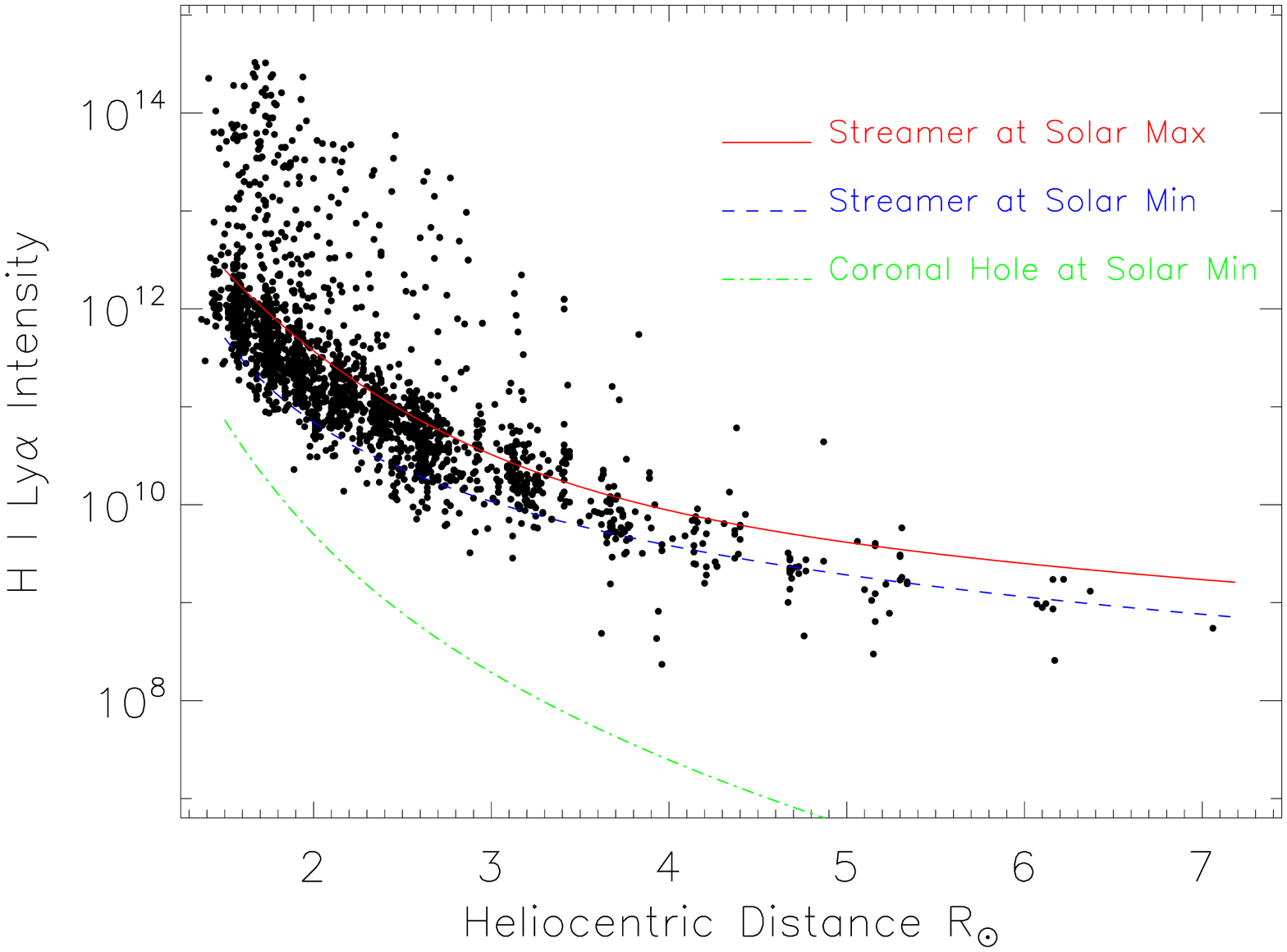} 
\includegraphics[width=6.25cm,angle=90]{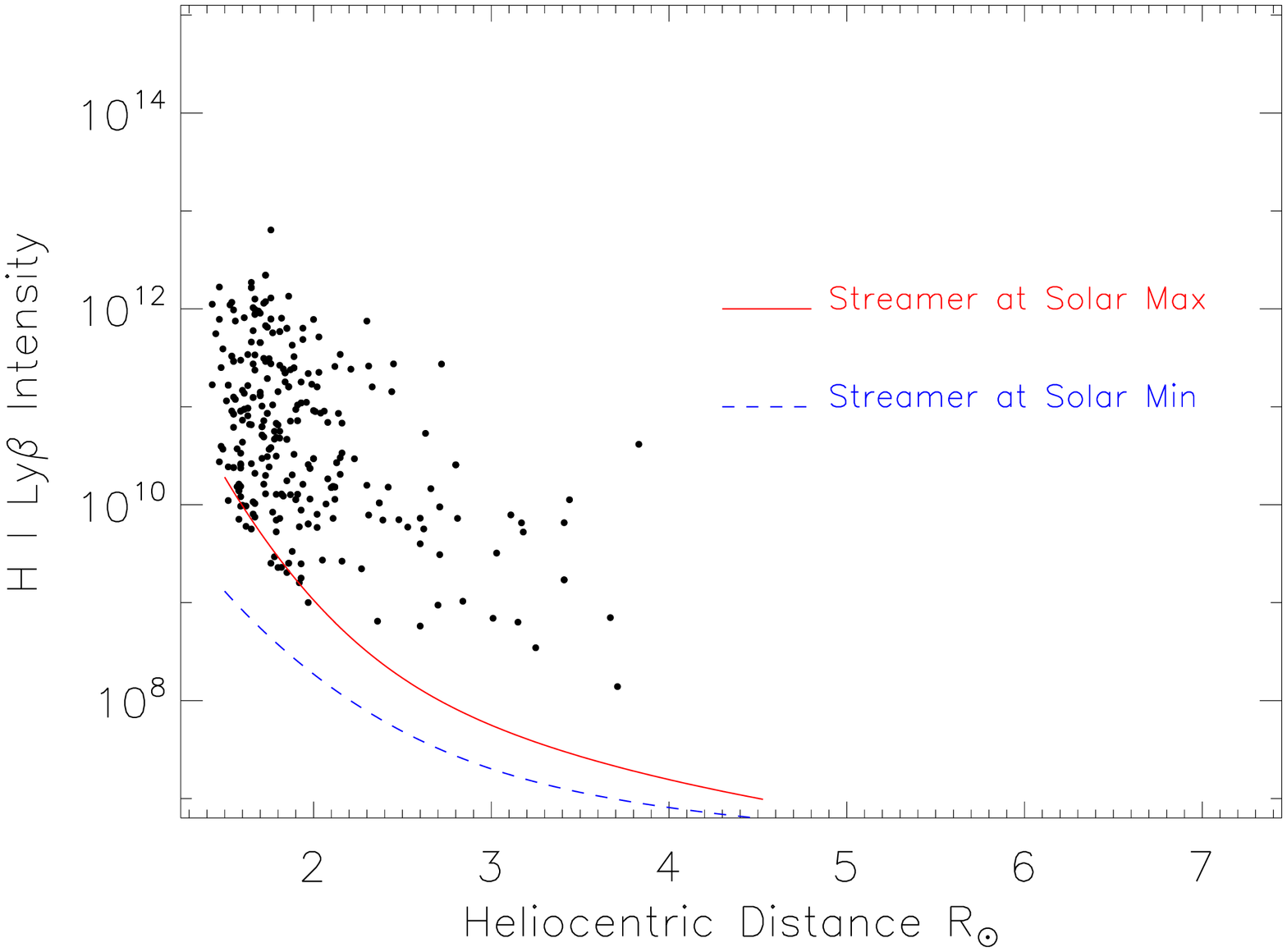}
\caption{CME intensity in H~I Ly$\alpha$ (left panel) and Ly$\beta$ (right panel) as a function of the heliocentric heights. 
		For comparison intensities measured in streamer at solar maximum and minimum and in coronal hole are also plotted.} 
\label{cme_ly}
\end{figure}

\begin{figure}
\centering
\includegraphics[width=6.25cm,angle=90]{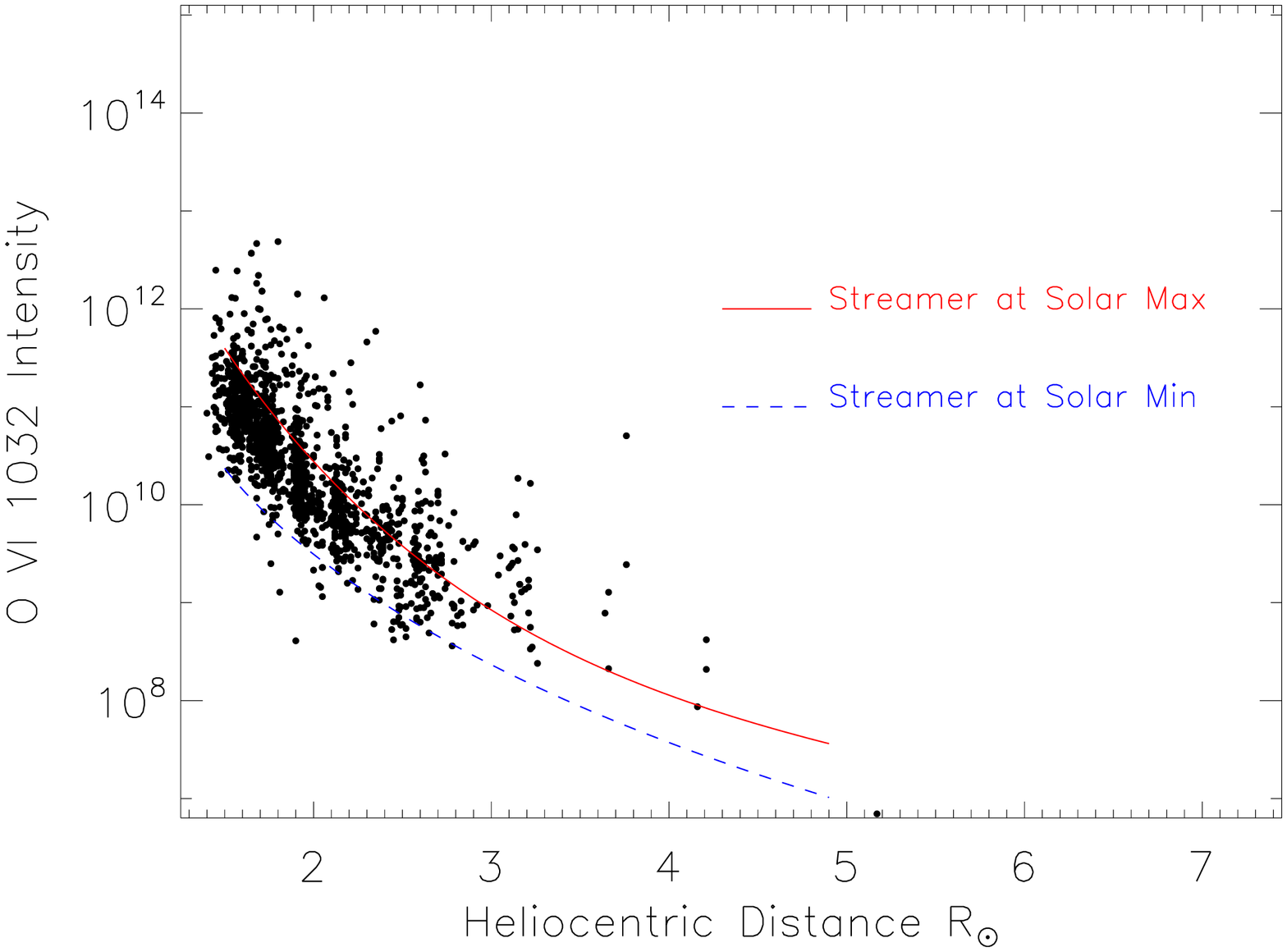} 
\includegraphics[width=6.25cm,angle=90]{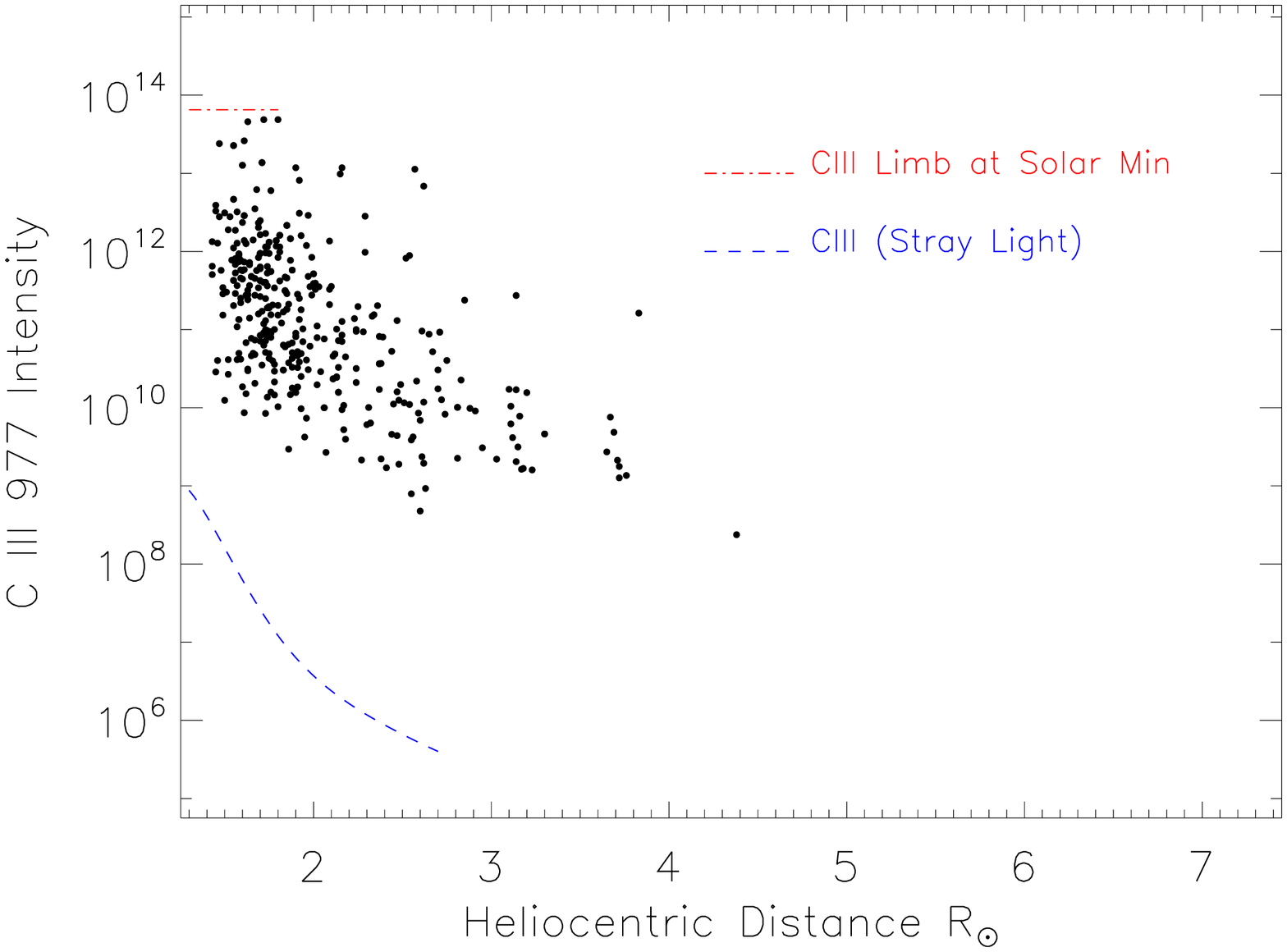}
\caption{CME intensity in O~VI 1032\AA\/ (left panel) and C~III 977\AA\/ (right panel) as a function of the heliocentric heights. 
	For comparison with O~VI the intensities measured in streamer at solar maximum and minimum are also plotted. 
	For the C~III lines the stray light value and the limb intensity are plotted.}
\label{cme_o6c3}
\end{figure}

We studied the spectral line intensity distributions by considering for each event only the heights where the CME has been detected 
and selecting for each height the maximum observed intensity.
The H~I Ly$\alpha$ and Ly$\beta$ line intensity distributions of the UVCS CMEs as functions of height are plotted in the left and right 
panels of Figure~\ref{cme_ly}, respectively.  As comparison the typical intensities in streamers at solar minimum and maximum and in 
coronal hole are also plotted.  These reference values are determined from the UVCS intensity Carrington maps used for coronal rotation 
studies \citep{gio08, man11b,man12b}.
The O~VI 1032\AA\/ (left panel) and C~III 977\AA\/ (right panel) line intensities in the UVCS CMEs are shown in Figure~\ref{cme_o6c3}.   
The O~VI intensities are compared to the streamer intensities at solar minimum and maximum, while for the C~III line, 
which is not a coronal line, the values for stray light \citep{Cra10} and the limb values are plotted. 
We note that the C~III in CMEs, within 2~R$_{\odot}$\ can reach intensity as high as the value observed at the solar limb.

At almost all the heights the H~I Ly$\alpha$ and Ly$\beta$ intensities in CMEs can reach values more than two orders of magnitude 
higher than the values detected in solar maximum streamers at the same heights.  
The O~VI intensity increase in CME with respect to coronal streamer is slightly less pronounced, but still about one order of magnitude or more.  
A few events at the lower heights approach the quiet Sun disk intensities of 
$5.2\times 10^{15}$, $4.1\times 10^{13}$, and $1.9\times 10^{13}$~photons cm$^{-2}$~s$^{-1}$~sr$^{-1}$ in those three lines, respectively.  
Line intensities decrease with height as the ejected material moves outward and expands, decreasing its density and therefore its emission. 
In the brightest events, the plotted intensities are underestimated, because the detector deadtime correction is probably significant 
\citep{Wil02}.  
Moreover, in the case of fast CMEs the integration time is longer than the time spent by the dense plasma to cross the instrumental slit, 
therefore, the measured intensity could be reduced by the fraction of the exposure time when CME plasma was in the slit.

Thus, we infer that for brightest events the UV lines intensity values determined by UVCS could be a lower limit of the true values.
However the majority of the events observed in H~I Ly$\alpha$  and O~VI have intensity below the typical solar maximum streamer, 
therefore they are detected in intensity only by running difference time series.

Comparison of H~I Ly$\alpha$ with C~III and O~VI lines indicates that the brightest H~I Ly$\alpha$ events are the ones with very 
low temperatures, and this is often corroborated by narrow H~I Ly$\alpha$ line profiles.  
The high intensities are observed in small regions along the UVCS slit, indicating that high densities are also required.  
Both of these properties indicate that the very bright material is prominence ejecta. 
   
\subsection{Interpretation of UV Spectra}\label{interpretation}

We examined the UVCS CME spectra for signatures of specific CME features,
such as shock waves and current sheets. Because of the large number of
events these classifications are not definitive, but they provide a
starting place for more detailed analysis and an approximate
idea of the frequency with which these features are detected.  In many
cases a question mark indicates that there was some hint of a feature, but
that more analysis is needed.

We identified the CME front, or leading edge, when the coronal lines brightened along much of the length of the UVCS slit 
at a time corresponding to the passage of the CME front seen in LASCO.  
Since the slit was often below the height of the LASCO occulter, this required some estimation. 
The CME front was clearly identified in about 
15\% 
of the events, and 
possibly identified in another 17\%. 

The CME void, or cavity, was identified as a dimmer region inside the leading edge, corresponding to the dark area seen in 
LASCO difference images.  It is difficult to be certain about projection effects due to emission from the front and back sides 
of the CME bubble.  Diffuse high temperature emission was sometimes seen and taken as a signature of emission from the void. 
For about 
13\% 
of the events the CME void has been identified 
, and for another 12\%  it is
questionable.

The shock waves observed by UVCS have been characterized by larger line widths than the background corona in the 
O~VI doublet due to shock heating \citep{Ray00,Man02,Rao04,Cia05,Cia06,Bem10,Man11}.
The emission tends to be faint, and the signature of broad O~VI lines requires poor ion-ion thermal equilibration, 
so there may be other shocks that we failed to identify. 
Of the 1059 events observed by UVCS, for 156 events there was  a type II radio burst, which could be associated with 
the CME because it has been detected in the same time window.  
In most cases, UVCS did not detect the shock wave either because the shock formed at a height above the 
UVCS observation, because UVCS caught only a portion of the CME that did not include the shock, or because 
the shock emission was too faint compared to foreground and background emission.
Shock waves have been identified for 16 events observed by UVCS, of these only 10 are associated with a type II radio burst. 
Moreover, there are 83 possible detections of shock fronts and more detailed analysis is necessary.

UVCS observations of current sheets are generally characterized by a narrow feature along the UVCS slit that is bright in 
high temperature lines such as Fe~XVIII $\lambda$975 \citep{Cia02,Ko03,Bem06,Cia08}, though \citet{Lin05} identified
a narrow feature seen in LASCO but faint in the UV lines as a current sheet.  
In many cases, the Fe~XVIII line was outside the observed wavelength range, and bright, narrow features in Si XII emission are 
identified as current sheets (e.g., \citet{Aur09}).  
If there are cool current sheets, they would generally not be identified. 
Only 
24
events clearly show CS signature, but others 
53 
are questionable.

We identify cool, bright material as prominence ejecta.  The presence of low ionization states such as C~III, N~III and O~V is 
the most common indicator. About 38\% of CMEs detected show clear signature of prominence material, while other 33\% of events 
show possible prominence material. In some cases, bright H~I~Ly$\alpha$ with a small line width indicates an ion temperature
far below coronal temperatures.

In some events, UVCS detects O~VI photons that are emitted from the flare and resonantly scattered from O~VI ions in 
the corona \citep{Ray07}.  This can be used to measure the transition region emission during the impulsive phase of the flare.  
In UVCS data, this emission is seen as a simultaneous O~VI brightening all along the slit.  It vanishes when the CME reaches 
the slit, because O~VI ions within the CME are Doppler shifted away from the low velocity emission of the flare.  
The intensity ratio of O~VI 1032 \AA\/ to 1037 \AA\/ is 4:1, because the intensity ratio of the flare emission is 2:1 and 
the scattering cross section of the 1032 \AA\/ line is twice that of the 1037 \AA\/ line.  
\citet{Joh11} have made a more detailed analysis of the events listed in the catalog and derived transition region luminosities for 29 events.  
There is a strong bias towards events quite close to the solar limb, since the intensity of the flare emission in the corona drops 
rapidly with distance from the flare.

The leg of the CME is taken to be the flank of the CME, perhaps consisting
of coronal material that is displaced and compressed transversely rather than radially.
It is often seen as brightening of ions seen at normal coronal temperatures
that appears at one end of the UVCS slit when the position angle of UVCS
does not place it directly above the eruption site.  The brightening can
move toward the center of the slit, and in some cases it moves back later in the event as the corona relaxes 
after the CME \citep{Cia03}.
For about 
13\% 
of the events the CME leg has been identified 
, and for another 16\%  it is
questionable.

In some CMEs, a narrow feature is seen whose pattern of motion along the 
slit and Doppler shift indicates helical motion \citep{Cia00}.  This is
likely to be the twisting or untwisting of a filament, perhaps a magnetic
flux tube connecting the solar surface to the CME core.  These features
are generally seen in cool ions, probably because most of the bright,
high contrast emission comes from low temperature plasma. 
We identified helical structures in 18 events and there are 
another
59 possible helical structures.

The identifications of the features described above are summarized in Table 3.  
The small number of shock waves and the large number of prominence detections reflect the sensitivity of UVCS to bright,
localized emission compared to faint, diffuse emission.  
The small number of definite current sheets compared to the large number of possible ones may reflect the fact that 
relatively few observations included the Fe~XVIII line.  
Further analysis, including comparison with white-light signatures of current sheets and longer time intervals after the CME, 
has greatly increased the number of current sheets 
detected both in WL images and UV spectra 
\citep{Cia12}.
The total number of these events is
78, among which 28 show emission in Fe~XVIII line, and the others are detected in cooler lines.
 
Of the 1059 CMEs of the UVCS catalog, 646 were associated with an X-ray flare which occurs in a narrow time window 
around the CME onset, but, the NOAA event list provides the coordinates on solar disk of only 392 of these flares.
Comparing the flares and CME positions we verified that 281 X-ray flares are also spatially associated with the CME detected by UVCS.
In some cases a flare occurred behind the limb,  and since UVCS is most sensitive to eruptions near the limb, a significant
fraction should be backside events. 
Over 281 events with clear association between CME and X-ray
flare 76 (23\%) show a red-shift, and 98 a blue-shift (35\%),
while over 413 without X-ray
flare 95 (23\%) show a red-shift and 74 (18\%) a blue-shift.
A more detailed work on single events reported in our catalog can help to extend the understanding of CME association with 
solar
flares, 
as previously studied by \citet{Ray03}
for three events associated with X-class flares.

\section{Summary}

We have made a systematic search for CMEs in UV spectra obtained with UVCS from 1996 through 2005.
The UVCS CME catalog contains 1059 events, or about on tenth the number of white-light CMEs in that period.  
The statistical properties of the UVCS and LASCO events are similar except that
UVCS tends to detect the more energetic, more massive and wider events.
The catalog includes a preliminary categorization of the types of
features observed, including shock fronts, current sheets and prominence
material.
Spectral lines covering the range log$_{10}$(T/K)=4.5 to log$_{10}$(T/K)=6.9 are
detected in the UVCS spectra.  In a few of the brightest events,
the intensities approach the solar disk values.  While Doppler shifts 
above 1000~km~s$^{-1}$ are sometimes seen, the average Doppler shifts
are about 1/10 the speeds given in the LASCO catalog.

The UVCS CME catalog is available to the public on World Wide Web, maintained and revised to include potentially missing 
events and catalog users suggestions.
The present release in also embedded in the online LASCO CME catalog.
The data in the UVCS CME catalog will make it easy for anyone investigating a particular CME to determine whether there is 
useful UVCS data on that event.
It also provides a means to find events for which UVCS data can provide especially strong constraints on theoretical models.
For detailed analysis and investigation we urge the user to get the original data, directly linked and downloadable from the 
single event web page.

\begin{acknowledgments}
This work was supported by 
National Aeronautics and Space Administration (NASA)
grants 
NNG06GG78G, NNX09AB17G and NNX11AB61G to the Smithsonian Astrophysical Observatory.
The LASCO CME catalog is generated and maintained at the CDAW Data Center by NASA and 
the Catholic University of America in cooperation with the Naval Research Laboratory. 
SOHO is a project of international cooperation between the European Space Agency (ESA) and the NASA. 
UVCS is a joint project of NASA, Italian Space Agency (ASI) and the Swiss Funding Agencies.

\end{acknowledgments}

%
%
%
%
%
%
%



\end{article}

\clearpage

\begin{table}
\caption{UVCS CMEs not listed in CME LASCO Catalog \label{uvcsonly}}
\centering
\begin{tabular}{c c l }
\hline
Date  	&	UVCS Roll & Comments \\
\hline
1997/08/23 & 270   	& CME at 08/23 23:15UT at 290$^{\circ}$ not in LASCO catalog \tablenotemark{a} 	\\
1997/09/30 & 225   	& CME at 10/01 01:29UT at 250$^{\circ}$ not in LASCO catalog \tablenotemark{a} 	\\
1998/03/31 & ~60	& No evidence of transient in LASCO data \tablenotemark{b} 					\\
1998/05/03 & 180	& No evidence of transient in LASCO data  \tablenotemark{b}					\\
1998/11/22 & 270	& No LASCO data from 11/12 16:18UT to 11/23 21:36UT, 1998 \tablenotemark{c} 	\\
1998/11/23 & ~90	& No LASCO data from 11/12 16:18UT to 11/23 21:36UT, 1998 \tablenotemark{c} 	\\
1998/11/23 & 225	& No LASCO data from 11/12 16:18UT to 11/23 21:36UT, 1998 \tablenotemark{c} 	\\
1999/04/23 & 250   	& CME/jet at 04/23 22:30UT at 230$^{\circ}$ not in LASCO catalog \tablenotemark{a}~ \tablenotemark{d}	\\
1999/05/09 & ~90	& CME/jet at 05/09 19:28UT at 95$^{\circ}$ not in LASCO catalog \tablenotemark{a}~ \tablenotemark{d} 	\\
1999/05/25 & 270	& CME at 05/25 01:26UT at 270$^{\circ}$ not in LASCO catalog \tablenotemark{a} 					\\
1999/06/27 & 330	& No evidence of transient in LASCO data \tablenotemark{b}									\\
1999/07/07 & 135	& Narrow blob, LASCO narrow bright feature 12:30UT, 135$^{\circ}$ \tablenotemark{a}~ \tablenotemark{d} \\
1999/07/11 & 270	& No evidence of transient in LASCO data (possible active streamer) \\
		   &		& narrow blob in UVCS \tablenotemark{b} \\
1999/08/15 & ~51	& jet at 08/15 15:54UT at 45$^{\circ}$ not in LASCO catalog \tablenotemark{a}~ \tablenotemark{d} 		\\
1999/08/26 & 292	& jet \citep{ko05} \tablenotemark{d} 												\\
1999/09/11 & 330	& part of the extended outflows after the CME at NNW on 09/10 						\\
		   &		& may correspond to a narrow limited-length jet-like material seen in C2 \tablenotemark{d}	\\
\hline
\end{tabular}
\tablenotetext{a}{7 events. May be included in CDAW CME catalog.}
\tablenotetext{b}{4 events. 
Not clearly related to transient in LASCO data.}
\tablenotetext{c}{3 events. No LASCO data available (Nov 1998).}
\tablenotetext{d}{5 events. Jet or jet-like features.}
\end{table}

\begin{table}
\caption{CME Spectral Lines observed by UVCS \label{tablines}}
\centering
\begin{tabular}{l l c l}
\hline
Ion 		& Wavelengths~(\AA)					& log$_{10}$(T/K)		& Comments \\
\hline
H I		& 1215.67, 1025.57, 972.54, 949.97		& 4.5		&   \\
He II   	& 1084.98                            				& 4.9		&   \\
C II    	& 1036.34, 1037.02                   			& 4.6		&   \\
C III   	&  977.02, 1174.93                   			& 4.8		&   \\
N II    	& 1084.56                            				& 4.6		&   \\
N III   	& 989.79, 991.58                     			& 4.8		&   \\
N V     	& 1238.82,1242.80                    			& 5.2		&    \\
O III     	& 597.82                    					& 5.0		&  \\
O V     	& 629.73                    					& 5.3		&  \\
O V     	& 1213.85, 1218.39                   			& 5.3		& Density diagnostic \\
O VI    	& 1031.93, 1037.61                   			& 5.4		& Density diagnostic \\
Ne VI   	& 999.18, 1005.69                         		& 5.6		& \\
Mg X    	& 609.79, 624.93                     			& 6.8		&  \\
Al XI     	& 550.03                    					& 6.9		&  \\
Si III  	& 1206.50, 1303.32                   			& 4.7		& Temperature diagnostic \\
Si VIII 	& 944.47, 949.35                       			& 5.9		& \\
Si IX   	& 950.17                             				& 6.1		& \\
Si XII 	& 499.40, 520.67                     			& 6.9		& \\
S V   	& 1199.14                             				& 5.2		&  \\
S VI		& 944.52                             				& 5.3		& \\
S X		& 1196.25                             			& 6.2 	& \\
Ca XIV  	& 943.59                              				& 6.7		& Current Sheet\\
Fe X  	& 1028.02							& 6.1		&  \\
Fe XII  	& 1242.00                            				& 6.2		&  \\
Fe XIII  	& 487.06                            				& 6.2		&  \\
Fe XVIII	& 974.86                             				& 6.9		&  Current Sheet \\
\hline
\end{tabular}
\end{table}

\begin{table}
\caption{UVCS CMEs Interpretation \label{tabint}}
\centering
\begin{tabular}{l r r r}
\hline
Feature 		& Yes 	& No 	& ?~ \\
\hline
Front		& 163 	& 719 	& 177	\\
Void			& 140 	& 797 	& 122	\\
Shock		& 16  	& 960 	&   83	\\
Current Sheet	& 24  	& 982 	&  53	\\
Prominence	& 406 	& 300 	& 353	\\
Flare O~VI	& 33  	& 993 	&   33	\\
Leg			& 129 	& 758 	& 172	\\
Helix			& 18  	& 982 	&   59	\\
\hline
\end{tabular}
\end{table}

\end{document}